\documentclass[
superscriptaddress,
groupedaddress,
reprint,
nofootinbib,
amsmath,amssymb,
aps,
prd,
floatfix,
]{revtex4-2}
\usepackage[utf8]{inputenc}
\usepackage{graphicx}
\usepackage{dcolumn}
\usepackage{bm}
\usepackage{hyperref}
\hypersetup{
    colorlinks=true,
    allcolors=blue,
}

\newcommand{\eq}[1]{Eq.~\eqref{#1}}

\DeclareMathOperator{\Ei}{Ei}
\DeclareMathOperator{\E}{E}

\usepackage{xcolor}

\usepackage[normalem]{ulem}
\usepackage{cancel}
\newif\ifrevision
\revisionfalse

\ifrevision
  \newcommand{\add}[1]{\textcolor{blue}{#1}}
  \newcommand{\del}[1]{\textcolor{red}{\sout{#1}}}

  \newcommand{\note}[1]{\textcolor{teal}{[Note: #1]}}
\else
  \newcommand{\add}[1]{#1}
  \newcommand{\del}[1]{}
  \newcommand{\note}[1]{}
\fi

\begin{document}

\title{Parameterized Post-Newtonian Analysis of Quadratic Gravity and Solar System Constraints}
\author{Jie Zhu}
 \email{jiezhu@cqu.edu.cn}
 \affiliation{Department of Physics and Chongqing Key Laboratory for Strongly Coupled Physics, Chongqing University, Chongqing 401331, P.R. China}

\author{Hao Li}
  \email{Corresponding author: haolee@cqu.edu.cn}
   \affiliation{Department of Physics and Chongqing Key Laboratory for Strongly Coupled Physics, Chongqing University, Chongqing 401331, P.R. China}

\date{\today}

\begin{abstract}
This work systematically investigates the post-Newtonian behavior of general quadratic gravity in the weak-field regime. 
By extending the Einstein-Hilbert action to include quadratic curvature terms as $\mathcal{L}\propto R-\lambda C^2+\mu R^2$, the theory introduces two massive modes: a scalar mode and a ghost tensor mode. 
Using the post-Newtonian expansion method, we derive the explicit expressions for the metric for a general source up to 1.5PN order.
Furthermore, for a point-mass source, we extend the solution to 2PN order and evaluate the effective parameterized post-Newtonian parameters $\gamma(r)$ and $\beta(r)$. 
The results show that deviations from General Relativity are exponentially suppressed. 
The theory has the feature $\gamma(r)\equiv 1$ when $m_R=m_W$, and to ensure that gravity remains attractive, we have $m_W>m_R/4$.
At short distance scales, the Newtonian potential no longer exhibits a $1/r$ behavior, but instead displays a polynomial dependence on $r$.
The leading correction to $\beta(r)$ exhibiting a characteristic $\mathcal{O}(r \ln (r)e^{-mr})$ dependence.
Based on the Solar System experiments, we derive preliminary constraints on the theory's parameters: $m_R,m_W\gtrsim23\mathrm{~AU}^{-1}$, corresponding to $\lambda\lesssim2.1\times10^{19}\mathrm{~m}^2$ and $\mu\lesssim7.1\times10^{18}\mathrm{~m}^2$. 
This study provides a theoretical foundation for future tests of quadratic gravity using pulsar timing arrays, gravitational-wave observations, and laboratory-scale short-range gravity experiments.
\end{abstract}

\maketitle

\section{Introduction}

General Relativity (GR), formulated by Albert Einstein in 1915, revolutionized our fundamental understanding of the universe by describing gravity not as a force, but as the geometric curvature of spacetime. 
For more than a century, GR has remained the cornerstone of modern physics, passing an array of increasingly stringent experimental and observational tests with remarkable precision~\cite{Will:2014kxa, Turyshev:2008ur, Berti:2015itd}. 
From the early success of explaining the anomalous perihelion precession of Mercury and the deflection of light by the Sun, to the modern-era milestones such as the direct detection of gravitational waves by LIGO/Virgo~\cite{LIGOScientific:2016aoc} and the direct imaging of black hole event horizons by the Event Horizon Telescope~\cite{EventHorizonTelescope:2019dse}, GR has consistently aligned with experimental data across diverse scales. 
These achievements have established GR as the standard model of gravitation, providing the indispensable framework for our current understanding of cosmology and astrophysics.

Despite its profound success, GR is widely regarded as an effective field theory rather than a final description of gravitation. 
From a theoretical perspective, the Einstein-Hilbert action is perturbatively non-renormalizable, posing a fundamental challenge to the formulation of a consistent quantum theory of gravity. 
Furthermore, at cosmological scales, the phenomena of dark energy and dark matter, as well as the need for a robust mechanism for cosmic inflation, suggest that the gravitational interaction may deviate from the predictions of GR in high-curvature or infrared regimes.
Among various modifications, Quadratic Gravity—which supplements the Einstein-Hilbert action with terms quadratic in the curvature tensors ($R^2$, $R_{\mu\nu}R^{\mu\nu}$ and $R_{\mu\nu\rho\sigma}R^{\mu\nu\rho\sigma}$)—stands out as a compelling candidate. The primary motivations for exploring quadratic gravity include:
\begin{itemize}
\item Renormalizability: 
% It was famously demonstrated by Stelle~\cite{Stelle:1976gc,Stelle:1977ry} that the addition of quadratic curvature terms renders the theory power-counting renormalizable \lh{the theory renormalizable in the sense of BPHZ}, 
It was famously demonstrated by Stelle~\cite{Stelle:1976gc,Stelle:1977ry, Buchbinder:1992gdx} that the addition of quadratic curvature terms renders the theory renormalizable in the sense of BPHZ,
offering a potential path towards a UV-complete theory of gravity.

\item Effective Field Theory (EFT) Predictions: Higher-order curvature terms naturally arise as quantum corrections in the low-energy effective action of more fundamental theories~\cite{Donoghue:1994dn}, such as string theory~\cite{Callan:1985ia, Gross:1986mw} or loop quantum gravity~\cite{Ashtekar:2011ni}.

\item Cosmological Inflation: The $R+ \alpha R^2$ model, pioneered by Starobinsky~\cite{Starobinsky:1980te, Mukhanov:1989rq, Whitt:1984pd, Myrzakulov:2014hca, Elizalde:2017mrn}, remains one of the most successful inflationary models, showing remarkable agreement with the latest cosmic microwave background (CMB) data from the Planck mission~\cite{Planck:2018jri}.

\item Singularity Resolution: Quadratic terms can potentially mitigate or resolve the curvature singularities found in black holes and the early universe (Big Bang) by introducing new degrees of freedom that dominate at high energies~\cite{Kanti:1998jd, Asorey:2024oxw, Duplessis:2015xva, Berej:2006cc, Nojiri:2005sx, Bamba:2010wfw, Bamba:2008ut}.
\end{itemize}

In recent years, quadratic gravity has attracted significant attention as a robust framework for exploring physics beyond GR.
The foundational understanding of the classical properties of quadratic gravity was significantly advanced by the seminal work of Stelle~\cite{Stelle:1977ry}. Within this framework, the inclusion of $R^2$, $R_{\mu\nu}R^{\mu\nu}$ and $R_{\mu\nu\rho\sigma}R^{\mu\nu\rho\sigma}$ terms in the action manifests as a modification of the gravitational interaction through the introduction of additional degrees of freedom: a massive scalar mode and a massive spin-2 ghost tensor mode. 
While quadratic gravity is often regarded as problematic due to the presence of ghost modes, recent analyses have demonstrated that it nevertheless satisfies classical causality constraints based on Shapiro time delay in shock-wave backgrounds~\cite{Edelstein:2021jyu, Edelstein:2024jzu}, indicating that such theories seem to provide viable descriptions of gravity.
Motivated by these foundational properties, extensive research has been conducted to explore the theory's impact on compact objects, like black hole solutions.
Black hole solutions beyond the Schwarzschild metric have been discovered~\cite{Lu:2015cqa}, and new families of static spherically symmetric solutions have been reported recently~\cite{Giacchini:2025mlv}.
Parallel to the study of static solutions, the dawn of gravitational-wave astronomy has sparked a rigorous effort to construct gravitational-wave waveforms within quadratic gravity. 
Because the additional massive modes contribute to the gravitational energy flux, they introduce specific modifications to the phase evolution of inspiralling binaries~\cite{Yagi:2011xp, Alves:2022yea, Alves:2025qcx}.
Despite these advancements in strong-field regimes, the precision tests within the Solar System remain a cornerstone for constraining modified gravity theories. The Parameterized Post-Newtonian (PPN)~\cite{Nordtvedt:1968qs, Thorne:1971iat, Will:1971zzb, Will:1971wt, Will:2014kxa, Will_1993, PoissonWill2014} formalism provides a powerful and model-independent framework to bridge the gap between theoretical predictions and weak-field observations. While the PPN parameters for certain subclasses of higher-order gravity, such as $f(R)$ theories, have been extensively investigated~\cite{Chiba:2006jp, Capozziello:2005bu, Capozziello:2007ms,  Capozziello:2008fn}, a comprehensive PPN analysis that simultaneously accounts for both the massive scalar and massive spin-2 degrees of freedom in general quadratic gravity is still evolving. Given that these massive modes introduce Yukawa-type corrections to the gravitational potential, they inevitably induce deviations in the PPN parameters from their General Relativistic values.

In this work, we perform a systematic study of the PPN formalisms within the context of general quadratic gravity. By considering the combined effects of the $R^2$, $R_{\mu\nu}R^{\mu\nu}$ and $R_{\mu\nu\rho\sigma}R^{\mu\nu\rho\sigma}$ terms, we derive the explicit expressions for the PPN metric up to 1.5PN, and PPN parameters $\gamma$ and $\beta$ for a point-like source. Our analysis aims to clarify how the characteristic mass scales of the theory influence the standard solar system tests, thereby providing a robust set of constraints that complement the results obtained from black hole physics and gravitational-wave observations.

The rest of this paper is structured as follows.
In Sec.~\ref{sec:intro}, we introduce the basics of the quadratic gravity;
In Sec.~\ref{sec:PN}, we calculate the exact solution of the metric up to 1.5PN for general sources;
In Sec.~\ref{sec:2PN}, we calculate the 2PN result for a point-like source and the effective PPN parameters $\gamma$ and $\beta$;
In Sec.~\ref{sec:exp}, we perform the constraint from the Solar System observations; 
And Sec.~\ref{sec:end} is the summary and the outlook.

% About PPN, not all theories can have a PPN expansion, Vainshteinian Mechanism, dCS, Scalar-Tensor with potential, EGB, and EdGB.

\section{Basics of Quadratic Gravity}\label{sec:intro}
% About the form of Quadratic Gravity
In quadratic gravity, the fourth-order terms that can be introduced are constituted by the invariants
$$R^2,~R_{\mu\nu}R^{\mu\nu},~R_{\mu\nu\alpha\beta}R^{\mu\nu\alpha\beta}~\mathrm{and}~\square R.$$
However, among these four invariants, only 2 contribute to the field equations.
In fact, the term $\square R$ is an explicit surface term, while the terms $R_{\mu\nu}R^{\mu\nu}$ and $R_{\mu\nu\alpha\beta}R^{\mu\nu\alpha\beta}$ can be written in terms of the Gauss-Bonnet $G^{2}$ and Weyl $C^2$ invariants:
\begin{equation*}
\begin{aligned}&G^{2}\equiv R^{2}-4R_{\mu\nu}R^{\mu\nu}+R_{\mu\nu\alpha\beta}R^{\mu\nu\alpha\beta},\\&C^{2}\equiv C_{\mu\nu\alpha\beta}C^{\mu\nu\alpha\beta}=\frac{1}{3}R^{2}-2R_{\mu\nu}R^{\mu\nu}+R_{\mu\nu\alpha\beta}R^{\mu\nu\alpha\beta},
\end{aligned}
\end{equation*}
where $C$ is the Weyl tensor
\begin{equation*}
C_{\mu\nu\rho\sigma}
= R_{\mu\nu\rho\sigma}
- \left(
g_{\mu[\rho} R_{\sigma]\nu}
- g_{\nu[\rho} R_{\sigma]\mu}
\right)
+ \frac{1}{3} R\, g_{\mu[\rho} g_{\sigma]\nu},
\end{equation*}
and here \(T_{[\mu\nu]} \equiv \frac12 ( T_{\mu\nu} - T_{\nu\mu} )\).
Thus,
\begin{equation*}
\begin{aligned}
R_{\mu\nu}R^{\mu\nu}=\frac{1}{3}R^{2}-\frac{1}{2}G^{2}+\frac{1}{2}C^{2},\\
R_{\mu\nu\alpha\beta}R^{\mu\nu\alpha\beta}=\frac{1}{3}R^{2}-G^{2}+2C^{2}.
\end{aligned}
\end{equation*}
Additionally, in 4 dimensions, the term $G^2$ is a topological invariant and does not contribute to the field equations.
Therefore, without loss of generality, we consider the action as
\begin{equation}
    S=\frac{1}{2\kappa}\int dx^4 \sqrt{-g}\left(R-\lambda C^2+\mu R^2\right)+S_m, \label{eq:action}
\end{equation}
where $\kappa=8\pi G$, and $S_m$ is the action for matter.
The action~(\ref{eq:action}) introduces a massive spin-2 tensor mode with mass as $m_W^2=\frac{1}{2\lambda}$, and a massive scalar mode with mass $m_R^2=\frac{1}{6\mu}$~\cite{Stelle:1977ry, Alves:2022yea, Alves:2025qcx}, and we will see it in the PN calculations.
In the dimension-4 spacetime, with the help of the Bach tensor \(B_{\mu\nu}\equiv(\nabla^{\rho}\nabla^{\sigma}+\frac{1}{2}R^{\rho\sigma})C_{\mu\rho\nu\sigma}\), the equations of motion from the action \eq{eq:action} is~\cite{Lu:2015cqa}
\begin{equation}
    \begin{aligned}
    R_{\mu\nu}&-\frac{1}{2}Rg_{\mu\nu}-4\lambda B_{\mu\nu}+2\mu R\left(R_{\mu\nu}-\frac{1}{4}Rg_{\mu\nu}\right)\\
    &+2\mu\left(g_{\mu\nu}\square R-\nabla_{\mu}\nabla_{\nu}R\right)=\kappa T_{\mu\nu}.\label{eq:eom1}
    \end{aligned}
\end{equation}
In addition, since the Bach tensor is traceless, the trace of the field equation \eq{eq:eom1} is
\begin{equation}
    6\mu\square R-R=\kappa T,
\end{equation}
% \lh{since the Bach tensor is traceless.}
Combining the two equations, we get
\begin{equation}
\begin{aligned}
&R_{\mu\nu}-4\lambda B_{\mu\nu}+2\mu R\left(R_{\mu\nu}-\frac{1}{4}Rg_{\mu\nu}\right)\\
&-2\mu\left(\frac{1}{2}g_{\mu\nu}\square R + \nabla_{\mu}\nabla_{\nu}R\right)=\kappa\left(T_{\mu\nu}-\frac{1}{2}g_{\mu\nu}T\right).\label{eq:eom2}
\end{aligned}
\end{equation}
In the following, we will calculate the PN expansion of the quadratic gravity from \eq{eq:eom2}.

\section{Exact PN solutions up to 1.5PN}\label{sec:PN}
The PPN~\cite{Nordtvedt:1968qs, Thorne:1971iat, Will:1971zzb, Will:1971wt, Will:2014kxa, Will_1993, PoissonWill2014} formalism provides a powerful and model-independent framework to bridge the gap between theoretical predictions and weak-field observations. 
However, the PPN framework is not universally applicable to all classes of modified gravity theories, and in certain cases, one must introduce novel potentials that fall outside the scope of the traditional PPN framework.
For instance, in gravity theories with parity violation, the metric component $g_{0i}$ requires the introduction of a new potential to capture the parity violation~\cite{Qiao:2021fwi}.
Another example is found in the regularized 4D Einstein-Gauss-Bonnet theory~\cite{Toniato:2024gtx} and the scalarized Einstein-Gauss-Bonnet theories ~\cite{Richarte:2025dag}, where the 2PN equations of motion require a novel potential $\Phi_\mathcal{G}$ to fully capture the higher-curvature corrections.
Such phenomena are particularly common in gravitational frameworks equipped with the Vainshtein screening mechanism~\cite{Vainshtein:1972sx}, where non-linear interactions dominate.
In order to account for the modifications induced by the Vainshtein screening, the Parametrized Post-Newtonian-Vainshteinian framework was developed~\cite{Avilez-Lopez:2015dja}, necessitating the inclusion of an infinite hierarchy of intricate new potentials beyond the standard PPN set.
In the context of quadratic gravity, we shall demonstrate that the presence of the two additional massive modes necessitates a significant extension of the gravitational potential set. 
Specifically, at the 1PN order, it is required to introduce two new potentials to account for the massive scalar and spin-2 contributions. 
At the 1.5PN level, at least one additional potential must be incorporated. However, at the 2PN order, the number of required potentials undergoes a combinatorial explosion, rendering the standard PPN potential approach increasingly intractable. 
Consequently, to maintain analytical clarity and physical insight, we shift our focus toward a point-source analysis to derive the higher-order corrections.

% We now expand the field equation~(\ref{eq:eom2}) as displayed in the preceding section up to the second post-Newtonian order.
Following the setup in the previous section, we now expand the field equations (\ref{eq:eom2}) to the second post-Newtonian order within the PPN formalism.
Assuming that the gravitating source matter is constituted by a perfect fluid that obeys the post-Newtonian hydrodynamics, we can write the energy-momentum tensor as
\begin{equation}
T^{\mu\nu}=\begin{pmatrix}\rho+\rho\Pi+p\end{pmatrix}u^{\mu}u^{\nu}+pg^{\mu\nu},\label{eq:emt}
\end{equation}
where $\rho$ is the rest energy density, $\Pi$ is the specific internal energy, $p$ is the pressure, $u^\mu=u^0 (1,v^i)$ is the four-velocity, and its three-velocity is $v^i$.
The energy-momentum tensor~(\ref{eq:emt}) is adequate to derive the post-Newtonian expansion of the gravitational field surrounding a fluid body, such as the Sun, or within a compact binary system~\cite{PoissonWill2014}.
When expanding the energy-momentum tensor in powers of $v/c$, we find that the equation of state and the internal energy are of order two, while the time derivative is of order one; specifically, $p/\rho\sim\Pi\sim(v/c)^{2}$ and $\partial_t\sim(v/c)$~\cite{PoissonWill2014}. 
Thus, the energy-momentum tensor~(\ref{eq:emt}) can be expanded in the form
\begin{equation}
\begin{aligned}&T_{00}=\rho\left(1+\Pi+v^{2}-h_{00}^{(2)}\right)+\mathcal{O}(6),\\&T_{0j}=-\rho v_{j}+\mathcal{O}(5),\\&T_{ij}=\rho v_{i}v_{j}+p\delta_{ij}+{\mathcal{O}}(6).\end{aligned}
\end{equation}

With respect to the metric $g_{\mu\nu}$, we expand it around a flat Minkowski background,
\begin{equation}
\begin{aligned}
g_{\mu\nu}&=\eta_{\mu\nu}+h_{\mu\nu}\\
&=\eta_{\mu\nu}+h_{\mu\nu}^{(1)}+h_{\mu\nu}^{(2)}+h_{\mu\nu}^{(3)}+h_{\mu\nu}^{(4)}+\mathcal{O}(5).
\end{aligned}
\end{equation}
The generic expansion of the metric components read~\cite{PoissonWill2014}
\begin{equation}
\begin{aligned}
&g_{00}=-1+h_{00}^{(2)}+h_{00}^{(4)}+\mathcal{O}(6),\\
&g_{0i}=h_{0i}^{(3)}+h_{0i}^{(5)}+\mathcal{O}(5),\\
&g_{ij}=\delta_{ij}+h_{ij}^{(2)}+\mathcal{O}(4).
\end{aligned}
\end{equation}
To get the post-Newtonian expansion of the field equation~(\ref{eq:eom2}), we use the \texttt{xPPN} package developed in Mathematica by Hohmann~\cite{Hohmann:2020muq}.

\subsection{Exact 1PN Solution}\label{sec:1PN}
Here we start by solving the second-order equations,
\begin{equation}
\begin{aligned}
&-\frac{\kappa \rho}{2}-\frac{1}{2}\Delta h_{00}^{(2)}+(\frac{2\lambda}{3}+\mu)\Delta\Delta h_{00}^{(2)}\\
&+(\frac{\lambda}{3}-\mu)\Delta \left(\Delta h_{aa}^{(2)}-\partial_a\partial_bh_{ab}^{(2)} \right)=0,\label{eq:2:1}
\end{aligned}
\end{equation}
\begin{equation}
\begin{aligned}
&-\frac{\kappa \rho}{2}\delta_{ab}+\delta_{ab}(\frac{\lambda}{3}-\mu)\Delta(\Delta h_{00}^{(2)}+\partial_c\partial_d h_{cd}^{(2)}-\Delta h_{cc}^{(2)})\\
&+\frac{1}{2}\partial_a\partial_b h_{00}^{(2)} -\frac{1}{2} \partial_a\partial_b h_{cc}^{(2)}-(\frac{\lambda}{3}+2\mu)\partial_a\partial_b\Delta h_{00}^{(2)}\\
&+(\frac{2}{3}\lambda-2\mu)\partial_a\partial_b \partial_c\partial_d h_{cd}^{(2)}+(\frac{\lambda}{3}+2\mu)\partial_a\partial_b\Delta h_{cc}^{(2)}\\
&+\frac{1}{2}(\partial_a\partial_c h_{bc}^{(2)}+\partial_b\partial_c h_{ac}^{(2)}-\Delta h_{ab}^{(2)})\\
&-\lambda \Delta (\partial_a\partial_c h_{bc}^{(2)}+\partial_b\partial_c h_{ac}^{(2)}-\Delta h_{ab}^{(2)})=0,\label{eq:2:2}
% &+\frac{1}{2}\partial_a\partial_b (h_{00}^{(2)}-h_{cc}^{(2)}-\frac{2}{3}(\lambda+6\mu)\Delta h_{00}^{(2)}\\
% &+\frac{4}{3}(\lambda-3\mu)\partial_c\partial_d h_{cd}^{(2)})
\end{aligned}
\end{equation}
% where the Einstein summation convention is implied \lh{summation with which metric? with all indices down is not Einstein summation, need more clarification}, 
where summation over repeated indices is implied,
and \(\Delta\equiv \partial_a\partial_a\) is the Laplacian operator.
To solve the equation, we need to fix the gauge.
We choose the isotropic spatial gauge, i.e., we propose that the solution to be
\begin{equation}
    h_{00}^{(2)}=U_1, \quad h_{ab}^{(2)}=\delta_{ab}U_2,\label{eq:U12def}
\end{equation}
where $U_1$ and $U_2$ are the functions to be solved.
Substitude \eq{eq:U12def} into the field equations~(\ref{eq:2:1}) and (\ref{eq:2:2}), we have
\begin{equation}
-\frac{\kappa\rho}{2}-\frac{1}{2}\Delta U_1+(\frac{2\lambda}{3}+\mu)\Delta\Delta U_1 +(\frac{2\lambda}{3}-2\mu)\Delta\Delta U_2=0,\label{eq:O2U1}
\end{equation}
and
\begin{equation}
\begin{aligned}
&\partial_a\partial_b\left(\frac{1}{2}U_1-\frac{1}{2}U_2-(\frac{\lambda}{3}+2\mu)\Delta U_1-(\frac{\lambda}{3}-4\mu)\Delta U_2\right)\\
&+\delta_{ab}\left(-\frac{1}{2}\Delta U_2+(\frac{\lambda}{3}-\mu)\Delta\Delta U_1+(\frac{\lambda}{3}+2\mu)\Delta\Delta U_2\right)\\
&-\frac{\kappa\rho}{2}\delta_{ab}=0.\label{eq:O2U2}
\end{aligned}
\end{equation}
Here we first force the two parts of \eq{eq:O2U2} to be zero, i.e.,
\begin{align}
&\frac{1}{2}U_1-\frac{1}{2}U_2-(\frac{\lambda}{3}+2\mu)\Delta U_1-(\frac{\lambda}{3}-4\mu)\Delta U_2=0,\label{eq:O2U3}\\ 
&-\frac{1}{2}\Delta U_2+(\frac{\lambda}{3}-\mu)\Delta\Delta U_1+(\frac{\lambda}{3}+2\mu)\Delta\Delta U_2-\frac{\kappa\rho}{2}=0,\label{eq:O2U4}
\end{align}
and we will show that there exists a solution satisfying all of the equations.
To solve the equations~(\ref{eq:O2U1}), (\ref{eq:O2U3}) and (\ref{eq:O2U4}), we use the Fourier transformation.
We use a tilde to denote the Fourier transform of a given function,  i.e.,
\begin{equation}
\begin{aligned}
\tilde{U}_1(\mathbf{k},t)&=\mathcal{F}[U_1(\mathbf{x},t)],\\
\tilde{U}_2(\mathbf{k},t)&=\mathcal{F}[U_2(\mathbf{x},t)], \\
\tilde{\rho}(\mathbf{k},t)&=\mathcal{F}[\rho(\mathbf{x},t)],   
\end{aligned}
\end{equation}
and the Fourier transformations of equations~(\ref{eq:O2U1}), (\ref{eq:O2U3}) and (\ref{eq:O2U4}) become
\begin{equation}
\begin{aligned}
&\frac{1}{6}k^2(3+2(2\lambda+3\mu)k^2)\tilde{U}_1+\frac{2}{3}(\lambda-3\mu)k^4\tilde{U}_2=\frac{\kappa}{2}\tilde{\rho},\\
&(3+2(\lambda+6\mu)k^2)\tilde{U}_1+(-3+2(\lambda-12\mu)k^2)\tilde{U}_2=0,\\
&\frac{1}{3}(\lambda-3\mu)k^4\tilde{U}_1+\frac{1}{6}k^2(3+2(\lambda+6\mu)k^2)\tilde{U}_2=\frac{\kappa}{2}\tilde{\rho}.
\end{aligned}
\end{equation}
With $m_W^2=\frac{1}{2\lambda}$ and $m_R^2=\frac{1}{6\mu}$,the solutions to the above equations are
\begin{equation}
\begin{aligned}
&\tilde{U}_1=\kappa\left(\frac{1}{k^2}+\frac{1}{3}\frac{1}{k^2+m_R^2}-\frac{4}{3}\frac{1}{k^2+m_W^2}\right)\tilde{\rho}(\mathbf{k}),\\
&\tilde{U}_1=\kappa\left(\frac{1}{k^2}-\frac{1}{3}\frac{1}{k^2+m_R^2}-\frac{2}{3}\frac{1}{k^2+m_W^2}\right)\tilde{\rho}(\mathbf{k}).\\
\end{aligned}
\end{equation}
Using the Fourier convolution theorem, and
\begin{equation}
    \mathcal{F}^{-1}\left[\frac{1}{k^2+m^2}\right]=\frac{1}{4\pi}\frac{e^{-mr}}{r},
\end{equation}
we have the solution in the position space as
\begin{equation}
\begin{aligned}
U_1(\mathbf{x},t)=\frac{\kappa}{4\pi}\left(U+\frac{1}{3} U_R-\frac{4}{3}U_W\right),\\
U_2(\mathbf{x},t)=\frac{\kappa}{4\pi}\left(U-\frac{1}{3} U_R-\frac{2}{3}U_W\right),\\\label{eq:solU12}
\end{aligned}
\end{equation}
where
\begin{equation}
\begin{aligned}
U(\mathbf{x},t)&\equiv\frac{1}{r}*\rho(\mathbf{x},t) =\int\frac{\rho(\mathbf{x}^{\prime},t)}{|\mathbf{x}-\mathbf{x}^{\prime}|}d^3x^{\prime},\\
U_R(\mathbf{x},t)&\equiv\frac{e^{-m_R r}}{r}*\rho(\mathbf{x},t)=\int\frac{e^{-m_R |\mathbf{x}-\mathbf{x}^{\prime}|}}{|\mathbf{x}-\mathbf{x}^{\prime}|}\rho(\mathbf{x}^{\prime},t)d^3x^{\prime},\\
U_W(\mathbf{x},t)&\equiv\frac{e^{-m_W r}}{r}*\rho(\mathbf{x},t)=\int\frac{e^{-m_W |\mathbf{x}-\mathbf{x}^{\prime}|}}{|\mathbf{x}-\mathbf{x}^{\prime}|}\rho(\mathbf{x}^{\prime},t)d^3x^{\prime},\label{eq:URW}
\end{aligned}
\end{equation}
and the symbol $*$ denotes the convolution.
These potentials have properties
\begin{equation}
\begin{aligned}
\Delta U &= -4\pi\rho,\\
(\Delta-m_R^2)U_R &= -4\pi\rho,\\
(\Delta-m_W^2)U_W &= -4\pi\rho.\label{eq:potential_relation}
\end{aligned}
\end{equation}
We can see that to describe the post-Newtonian behavior of the quadratic gravity, at $\mathcal{O}(2)$ order, we need to define two new potentials $U_R$ and $U_W$.
Finally, the solution to 1PN 
% \lh{1PN to be consistent?} 
is
\begin{equation}
\begin{aligned}
&h_{00}^{(2)}=\frac{\kappa}{4\pi}\left(U+\frac{1}{3} U_R-\frac{4}{3}U_W\right),\\
&h_{ab}^{(2)}=\frac{\kappa}{4\pi}\left(U-\frac{1}{3} U_R-\frac{2}{3}U_W\right)\delta_{ab}. \label{eq:solO2}
\end{aligned}
\end{equation}
In the limit $m_W \to \infty, ~m_R \to \infty$, the result reduces to the GR case.
For an N-body system consisting of point-like sources, the solution is 
\begin{equation}
\begin{aligned}
&h_{00}^{(2)}=\sum_A\frac{2G m_A}{r_A}\left(1+\frac{1}{3}e^{-m_R r_A}-\frac{4}{3}e^{-m_W r_A}\right),\\
&h_{ab}^{(2)}=\sum_A\frac{2G m_A}{r_A}\left(1-\frac{1}{3}e^{-m_R r_A}-\frac{2}{3}e^{-m_W r_A}\right)\delta_{ab}.\label{eq:Nbd1PN}
\end{aligned}
\end{equation}

\subsection{Exact 1.5PN solution}\label{sec:1.5PN}

The third-order equation is
\begin{equation}
\begin{aligned}
&(1-2\lambda \Delta)(\partial_a\partial_b-\delta_{ab}\Delta)h_{0b}^{(3)}\\
&+\frac{4}{3}(\lambda-3\mu)\Delta\partial_0\partial_a h_{00}^{(2)}+\frac{2}{3}(\lambda+6\mu)\Delta\partial_0\partial_a h_{bb}^{(2)}\\
&+\frac{4}{3}(\lambda-3\mu)\partial_0\partial_a\partial_b\partial_c h_{bc}^{(2)}-2\lambda \Delta\partial_0\partial_b h_{ab}^{(2)}\\
&+\partial_0\partial_b h_{ab}^{(2)}-\partial_0\partial_a h_{bb}^{(2)}+2\kappa\rho v_a=0. \label{eq:O3V1}
\end{aligned}
\end{equation}

Using the solution~(\ref{eq:solO2}) and the relation~(\ref{eq:potential_relation}), we can simplify \eq{eq:O3V1} as
\begin{equation}
\begin{aligned}
&(1-2\lambda \Delta)(\partial_a\partial_b-\delta_{ab}\Delta)h_{0b}^{(3)}\\
=&-2\kappa \rho v_a + \frac{\kappa}{2\pi}\partial_0\partial_a U.
\end{aligned}\label{eq:O3V2}
\end{equation}
\eq{eq:O3V2} shows that  only the Weyl sector contribution enters the 1.5PN.
 % \lh{with or without spacing between number and PN?} corrections.
When $\lambda=0$, \eq{eq:O3V2} becomes the 1.5PN equation of GR.
Using the result of GR, we have
\begin{equation}
(1-2\lambda \Delta)h_{0a}^{(3)} = -\frac{\kappa+C_0}{4\pi}V_a-\frac{\kappa-C_0}{4\pi}W_a,\label{eq:O3V3}
\end{equation}
where $C_0$ is a gauge parameter, and $V^a$ and $W^a$ are potential defined as
\begin{equation}
\begin{aligned}
V_{a}(\mathbf{x},t)&\equiv\int\frac{\rho(\mathbf{x}^{\prime},t)v_{a}(\mathbf{x}^{\prime},t)}{|\mathbf{x}-\mathbf{x}^{\prime}|}d^{3}x^{\prime},\\
W_a(\mathbf{x},t)&\equiv\int\rho(\mathbf{x}^{\prime},t)\frac{\left[\mathbf{v}(\mathbf{x}^{\prime},t)\cdot(\mathbf{x}-\mathbf{x}^{\prime})\right](x-x^{\prime})_a}{|\mathbf{x}-\mathbf{x}^{\prime}|^3}d^3x^{\prime}.
\end{aligned}
\end{equation}
Here we choose the gauge $C_0=\kappa$ to eliminate the potential $W_a$, and \eq{eq:O3V3} turns to be
\begin{equation}
(1-2\lambda \Delta)h_{0a}^{(3)} = -\frac{\kappa}{2\pi}V_a.\label{eq:O3V4}
\end{equation}
\eq{eq:O3V4} is the screened Poisson equation, and it can be solved by the Fourier transformation, similar to the method in Sec.~\ref{sec:1PN}.
The solution is
\begin{equation}
h_{0a}^{(3)} = -\frac{\kappa}{2\pi}\mathcal{V}_a, \label{eq:solO3}
\end{equation}
where $\mathcal{V}_a$ is a new potential defined as
\begin{equation}
\begin{aligned}
\mathcal{V}_a(\mathbf{x},t)&\equiv\frac{m_W^2}{4\pi} \frac{e^{-m_W r}}{r}*V_a(\mathbf{x},t)\\
&=\frac{m_W^2}{4\pi}\int\frac{e^{-m_W |\mathbf{x}-\mathbf{x}^{\prime}|}}{|\mathbf{x}-\mathbf{x}^{\prime}|}V_a(\mathbf{x}^{\prime},t)d^3x^{\prime}.\label{eq:newV}
\end{aligned}
\end{equation}
In the limit $m_W\to\infty$, the convolution kernel $\frac{m_W^2}{4\pi} \frac{e^{-m_W r}}{r}$ becomes $\delta^{(3)}(\mathbf{x})$, and the new potential $\mathcal{V}_a$ reduces to the GR case $V_a$.
If we keep the gauge parameter $C_0$, the solution is
\begin{equation}
h_{0a}^{(3)} = -\frac{\kappa+C_0}{4\pi}\mathcal{V}_a-\frac{\kappa-C_0}{4\pi}\mathcal{W}_a,
\end{equation}
where $\mathcal{W}_a$ is a new potential similar to $\mathcal{V}_a$ as
\begin{equation}
\begin{aligned}
\mathcal{W}_a(\mathbf{x},t)&\equiv\frac{m_W^2}{4\pi} \frac{e^{-m_W r}}{r}*W_a(\mathbf{x},t)\\
&=\frac{m_W^2}{4\pi}\int\frac{e^{-m_W |\mathbf{x}-\mathbf{x}^{\prime}|}}{|\mathbf{x}-\mathbf{x}^{\prime}|}W_a(\mathbf{x}^{\prime},t)d^3x^{\prime}.\label{eq:newW}
\end{aligned}
\end{equation}
The parameter $C_0$ is a pure gauge freedom at the 1.5PN level. It is kept so that the metric can be expressed in the standard PPN gauge and directly compared with the PPN parameters.
One may fix this gauge from the outset and still consistently perform the 2PN expansion, but the resulting expressions generally fall outside the standard PPN framework, although they remain physically equivalent.
Since the PN expansion of quadratic gravity already departs significantly from the standard PPN parameter framework, retaining the gauge parameter $C_0$ is no longer essential at this stage.

A noteworthy aspect of the 1.5PN corrections is that they are governed exclusively by the Weyl sector, implying that the massive spin-2 mode is the only active degree of freedom at this level.
This is also evident from the linearized treatment of quadratic gravity in Ref.~\cite{Alves:2022yea}: the massive scalar mode contributes to the diagonal components of the metric, whereas the massive spin-2 mode manifests as a contribution to the $g_{0i}$ sector.

For an N-body system consisting of point-like sources, the potential $V_a$ can be expressed as
\begin{equation}
V^a=\sum_A \frac{m_A v_A^a}{r_A}.
\end{equation}
To obtain $\mathcal{V}_a$ in the case of point-like sources, we need to compute the convolution of $1/r$ and $e^{-mr}/r$.
Denote $f=\frac{1}{r}*\frac{e^{-mr}}{r}$, a simpler way to perform the calculation is using the relation
\begin{equation}
(\Delta -m^2)f=-\frac{4\pi}{r}.\label{eq:f}
\end{equation}
Applying the Fourier transform, we get
\begin{equation}
-(k^2+m^2)\tilde{f}=-\frac{(4\pi)^2}{k^2}.
\end{equation}
So
\begin{equation}
\tilde{f}=\frac{(4\pi)^2}{m^2}\left(\frac{1}{k^2}-\frac{1}{k^2+m^2}\right),
\end{equation}
and the solution for $f$ is
\begin{equation}
f = \frac{4\pi}{m^2}\frac{1-e^{-mr}}{r}.\label{eq:fsol}
\end{equation}
Another way to solve \eq{eq:f} is that, since the result of $f$ should be spherically symmetric, \eq{eq:f} in spherical coordinate is
\begin{equation}
    f''(r)+\frac{2}{r}f(r)-m^2f(r)=-\frac{4\pi}{r}.
\end{equation}
% Choosing the boundary condition that $f(0)$ and $f(\infty)$ are finite, we obtain the same solution as \lh{in} \eq{eq:fsol}. 
Choosing the boundary condition that $f(0)$ and $f(\infty)$ are finite, we obtain the same solution as in \eq{eq:fsol}. 
This method, which reduces the convolution of two spherically symmetric expressions to solving an ordinary differential equation, is extensively used when computing the 2PN results.
Finally, for the N-body system consisting of point-like sources, we have
\begin{equation}
    \mathcal{V}_a =\sum_A \frac{m_A v_A^a}{r_A}(1-e^{-m_W r_A}),
\end{equation}
and
\begin{equation}
h_{0a}^{(3)} = -4G \sum_A \frac{m_A v_A^a}{r_A}(1-e^{-m_W r_A}).\label{eq:Nbd1.5PN}
\end{equation}

\section{2PN solution for a point-like source}\label{sec:2PN}

% With the exact solution \eq{eq:U12def} and \eq{eq:solO3}, 

With the the exact solution \eq{eq:U12def} and \eq{eq:solO3}, the fourth-order equations are 
\begin{widetext}

\begin{equation}
\begin{aligned}
&-\frac{1}{2}\Delta h_{00}^{(4)}
+\frac{1}{3}(-\lambda+3\mu)\partial_a\partial_b\Delta h_{ab}^{(4)}
+\frac{1}{3}(2\lambda+3\mu)\Delta\Delta h_{00}^{(4)}
+\frac{1}{3}(\lambda-3\mu)\Delta\Delta h_{aa}^{(4)}
+S_1=0,\label{eq:O4h1}
\end{aligned}
\end{equation}
\begin{equation}
\begin{aligned}
&\frac{1}{2}\partial_a\partial_b h_{00}^{(4)}
-\frac{1}{2}\partial_a\partial_b h_{cc}^{(4)}
+\frac{1}{2}\partial_c\partial_a h_{bc}^{(4)}
+\frac{1}{2}\partial_c \partial_b h_{ac}^{(4)}
-\frac{1}{2}\Delta h_{ab}^{(4)}
+\lambda \Delta \Delta h_{ab}^{(4)}
-\lambda \partial_a\partial_c\Delta h_{bc}^{(4)}
-\lambda  \partial_b\partial_c \Delta h_{ac}^{(4)}
\\&
-\frac{1}{3}(\lambda+6\mu)\partial_a\partial_b\Delta h_{00}^{(4)}
+\frac{1}{3}(\lambda+6\mu)\partial_a\partial_b\Delta h_{cc}^{(4)}
+\frac{2}{3}(\lambda-3\mu)\partial_a\partial_b\partial_c\partial_d h_{cd}^{(4)}
\\&
+\frac{1}{3}(\lambda-3\mu)\delta_{ab}\Delta\left(
\partial_c\partial_d h_{cd}^{(4)}
+\Delta h_{00}^{(4)}
-\Delta h_{cc}^{(4)}
\right)
+A_{ab}+\delta_{ab}B=0,\label{eq:O4h2}
\end{aligned}
\end{equation}
where $S_1$, $A_{ab}$, and $B$ are

% \begin{widetext}

\begin{equation}
\begin{aligned}
S_1=&
-\frac{1}{2}\kappa \rho\Pi-\frac{3}{2}\kappa p-\kappa \rho v_a v_a
+\partial_0\partial_a\mathcal{V}_a
-\frac{2}{3}(2\lambda+3\mu)\partial_0\partial_a\Delta \mathcal{V}_a
-\frac{3}{2}\partial_0\partial_0 U_2
-3\mu \partial_0\partial_0\Delta U_1
+9\mu \partial_0\partial_0\Delta U_2\\
&
+\frac{1}{2}U_2\Delta U_1
-\frac{1}{4}\partial_a U_1 \partial_a U_1
-\frac{1}{4}\partial_a U_2 \partial_a U_1
+\frac{1}{2}(4\lambda+3\mu)\partial_a \Delta U_1\partial_a U_1
+\frac{1}{6}(-4\lambda+21\mu)\partial_a \Delta U_2\partial_a U_1\\
&
-\frac{1}{3}(2\lambda+3\mu) \partial_a \Delta U_1\partial_a U_2
+\frac{10}{3}(-\lambda+3\mu)\partial_a \Delta U_2\partial_a U_2
+\frac{1}{12}(\lambda +6\mu)\Delta U_1\Delta U_1
-\frac{1}{6}(7\lambda +6\mu)\Delta U_1\Delta U_2\\ 
&
+\frac{1}{4}(-5\lambda +24\mu)\Delta U_2\Delta U_2
-\frac{2}{3}(2\lambda+3\mu)U_2\Delta\Delta U_1
+\frac{2}{3}(-\lambda+3\mu)U_1\Delta\Delta U_2
+2(-\lambda+3\mu)U_2\Delta\Delta U_2\\
&
+\frac{1}{12}(17\lambda+12\mu)\partial_a\partial_b U_1\partial_a\partial_b U_1
+\frac{1}{6}(\lambda+6\mu)\partial_a\partial_b U_1\partial_a\partial_b U_2
+\frac{1}{4}(-5\lambda+12\mu)\partial_a\partial_b U_2\partial_a\partial_b U_2,
\end{aligned}
\end{equation}

\begin{equation}
\begin{aligned}
A_{ab}=&
-\kappa \rho v_a v_b
% -\frac{1}{2}\partial_0(\partial_a \mathcal{V}_b+\partial_b\mathcal{V}_a)
+\frac{4}{3}(-\lambda+3\mu)\partial_0\partial_a\partial_b\partial_c\mathcal{V}_c
% +\lambda \partial_0\Delta(\partial_a \mathcal{V}_b+\partial_b\mathcal{V}_a)
-\frac{1}{2}(1-2\lambda\Delta)\partial_0(\partial_a \mathcal{V}_b+\partial_b\mathcal{V}_a)
+\partial_0\partial_0\partial_a\partial_b (\lambda U_1+(\lambda-6\mu) U_2)\\
&
+\frac{1}{4}\partial_a U_1 \partial_b U_1
-\frac{1}{4}\partial_a U_2 \partial_b U_1
-\frac{1}{4}\partial_a U_1 \partial_b U_2
+\frac{3}{4}\partial_a U_2 \partial_b U_2
+\frac{1}{6}(\lambda-12\mu)\partial_a\Delta U_1 \partial_b U_1
+\frac{1}{2}\lambda \partial_a\Delta U_2 \partial_b U_1\\
&
+\frac{1}{2}(\lambda+6\mu) \partial_a\Delta U_1 \partial_b U_2
+\frac{5}{6}(\lambda-12\mu) \partial_a\Delta U_2 \partial_b U_2
+\frac{1}{2}U_1\partial_a\partial_b U_1
+\frac{1}{2}U_2\partial_a\partial_b U_2
-\frac{1}{3}(\lambda+6\mu)U_1\partial_a\partial_b\Delta U_1\\
&
+\frac{1}{3}(\lambda+6\mu)U_2\partial_a\partial_b\Delta U_1
+\frac{2}{3}(\lambda-12\mu)U_2\partial_a\partial_b\Delta U_2
+\frac{1}{6}(\lambda-12\mu)\partial_a U_1 \partial_b \Delta U_1
+\frac{1}{2}(\lambda+6\mu)\partial_a U_2 \partial_b \Delta U_1
\\&
+\frac{1}{2}\lambda \partial_a U_1 \partial_b \Delta U_2
+\frac{5}{6}(\lambda-12\mu) \partial_a U_2 \partial_b \Delta U_2
-\frac{1}{3}(2\lambda+3\mu)\partial_a\partial_b U_1\Delta U_1
+\frac{1}{3}(-\lambda+3\mu)\partial_a\partial_b U_2\Delta U_1
\\&
-\frac{1}{3}(\lambda+6\mu)\partial_a\partial_b U_1\Delta U_2
-6\mu \partial_a\partial_b U_2\Delta U_2
-\frac{1}{6}(5\lambda+12\mu)\partial_a\partial_b\partial_c U_1\partial_c U_1
-\frac{1}{6}(\lambda+6\mu)\partial_a\partial_b\partial_c U_2\partial_c U_1
\\&
-\frac{1}{6}(\lambda+6\mu)\partial_a\partial_b\partial_c U_1\partial_c U_2
+\frac{1}{2}(\lambda-12\mu)\partial_a\partial_b\partial_c U_2\partial_c U_2
-\frac{1}{3}(\lambda+6\mu)\partial_b\partial_c U_1 \partial_a\partial_c U_1
\\&
+\frac{1}{3}(\lambda-3\mu)\partial_b\partial_c U_2 \partial_a\partial_c U_1
+(\lambda-6\mu)\partial_b\partial_c U_2 \partial_a\partial_c U_2
+\frac{1}{3}(\lambda-3\mu)\partial_b\partial_c U_1 \partial_a\partial_c U_2,
\end{aligned}
\end{equation}

\begin{equation}
\begin{aligned}
B=&
-\frac{1}{2}\kappa \rho \Pi+\frac{1}{2}\kappa p
-\frac{1}{2}\kappa \rho U_2
+\frac{1}{2}\partial_0\partial_0 U_2
+\frac{1}{2}U_2\Delta U_2
+\frac{1}{4}\partial_c U_1\partial_c U_2
+\frac{1}{4}\partial_c U_2\partial_c U_2
+\frac{2}{3}(-\lambda+3\mu)\partial_0\partial_c\Delta\mathcal{V}_c
\\&
+\frac{1}{3}(-\lambda+3\mu)\partial_0\partial_0\Delta U_1
-\frac{1}{3}(\lambda+15\mu)\partial_0\partial_0\Delta U_2
+\frac{5}{6}(\lambda-3\mu)\partial_c\Delta U_1\partial_c U_1
-\frac{1}{2}(\lambda+3\mu)\partial_c\Delta U_2\partial_c U_1
\\&
-\frac{1}{2}\lambda \partial_c\Delta U_1\partial_c U_2
-\frac{1}{6}(11\lambda+48\mu) \partial_c\Delta U_2\partial_c U_2
+\frac{1}{4}(\lambda-6\mu)\Delta U_1\Delta U_1
+\frac{1}{6}(-\lambda+12\mu)\Delta U_1\Delta U_2
\\&
-\frac{1}{12}(5\lambda+48\mu)\Delta U_1\Delta U_1
+\frac{1}{3}(\lambda -3\mu) U_1\Delta\Delta U_1
+\frac{1}{3}(-\lambda +3\mu) U_2\Delta\Delta U_1
-\frac{2}{3}(\lambda + 6\mu) U_2\Delta\Delta U_2
\\&
+\frac{1}{12}(7\lambda-12\mu) \partial_c\partial_d U_1 \partial_c\partial_d U_1
-\frac{1}{6}(\lambda+6\mu) \partial_c\partial_d U_2 \partial_c\partial_d U_1
-\frac{3}{4}(\lambda+4\mu) \partial_c\partial_d U_2 \partial_c\partial_d U_2.
\end{aligned}
\end{equation}

\end{widetext}

Using the Fourier transformation method shown in Sec.~\ref{sec:1PN} and~\ref{sec:1.5PN}, 
% we can maybe \lh{might be able to} obtain the 2PN results for general sources.
we might be able to obtain the 2PN results for general sources.
However, the complexity of the problem lies in the fact that we need to define a whole set of new potentials to express this result.
Using the solution for $U_1$ and $U_2$ as in \eq{eq:solU12}, 
and the properties \eq{eq:potential_relation}, we can reduce the term $\Delta U_1$ and $\Delta U_2$ into a linear combination of $U$, $U_R$, $U_W$ and $\rho$.
As a result, 
% Recall that in the 1PN calculation, we define two new potentials $U_R$ and $U_W$, and in the 1.5PN calculation, we define one new potential $\mathcal{V}_a$; these new potentials will enter the calculation of the 2PN result.
% More specifically, 
we need to calculate the convolution between the kernel $1/r$, $e^{-m_W r}/r$, and $e^{-m_R r}/r$, and the mixing between the potential $U$, $U_R$, and $U_W$ such as $UU_W$, $\partial_a U_W \partial_a U_R$, $\partial_a\partial_b U \partial_a\partial_b U_R$, etc.
For the 2PN results, we only need the $h_{00}^{(4)}$ part.
In the GR case or some modified gravity like Scalar-Tensor gravity, the equations for $h_{00}^{(4)}$ and $h_{ij}^{(4)}$ are decoupled, and we only need to solve the $tt-$ part of the fourth-order equations.
However, in quadratic gravity, the equations for $h_{00}^{(4)}$ and $h_{ij}^{(4)}$ are mixed, which brings more complexity in the 2PN result.
This can be easily seen from \eq{eq:O4h1} that in the limit $\lambda\to\infty$ and $\mu\to\infty$, \eq{eq:O4h1} reduces to
$-\frac{1}{2}\Delta h_{00}^{(4)}+S_1=0$, from which we can solve $h_{00}^{(4)}$ directly.
Although one may, in principle, write down a general expression for the 2PN result, the necessity of introducing and redefining a large number of new potentials renders the final expression exceedingly complicated. 
%As a consequence, such a result effectively loses its predictive power.\lh{???}

In order to investigate the 2PN phenomena of quartic gravity in the weak-field regime, we adopt the strategy of Ref.~\cite{Hohmann:2013rba}.
That is, we consider the gravitational field generated by a point-like mass $M$, which is described by the energy-momentum tensor~(\ref{eq:emt}) with
\begin{equation}
\rho=M\delta^{(3)}(\mathbf{x}),\quad \Pi=0, \quad p=0, \quad v_a=0.\label{eq:source}
\end{equation}
With this source, we can use the isotropic spherical coordinates, and the gravitational field can be expressed as
\begin{equation}
\begin{aligned}
g_{00}=&-1+2G_{\mathrm{eff}}(r)U_N(r)-2G_{\mathrm{eff}}^{2}(r)\beta(r)U_N^{2}(r)\\
&+\Phi^{(4)}(r)+\mathcal{O}(6),\\
g_{0j}=&{\mathcal{O}}(5),\\
g_{ij}=&\begin{bmatrix}{1+2G_{\mathrm{eff}}(r)\gamma(r)U_N(r)}\end{bmatrix}\delta_{ij}+\mathcal{O}(4),\end{aligned}
\end{equation}
where the Newtonian potential is defined as
\begin{equation}
     U_N(r)\equiv\frac{G M}{r},
\end{equation}
and $\Phi^{(4)}(r)$ are the terms of order $\mathcal{O}(4)$ which are not of the form $G_{\mathrm{eff}}^{2}(r)\beta(r)U^{2}(r)$, such as  the gravitational self-energy.
From the 1PN result in Sec.~\ref{sec:1PN}, we know that
\begin{equation}
\begin{aligned}
% \frac{G_{\mathrm{eff}}(r)}{G}
G_{\mathrm{eff}}(r)
&=1+\frac{1}{3}e^{-m_R r}-\frac{4}{3}e^{-m_W r},\\
\gamma(r)&=\frac{3-e^{-m_R r}-2e^{-m_W r}}{3+e^{-m_R r}-4e^{-m_W r}}. \label{eq:Gga}
\end{aligned}
\end{equation}
% In the limit $m_W\to\infty$, the result reduces to the result in $f(R)$ \lh{$f(R)\propto R^2$} gravity~\cite{Berry:2011pb}.
In the limit $m_W\to\infty$, the result reduces to the result in $f(R)$ gravity~\cite{Berry:2011pb}.
We can immediately see that if $m_R=m_W=m$, then $\gamma(r)=1$, reducing to the GR case.
In the following, we will calculate the effective 2PN parameter $\beta(r)$.

Since for spherical symmetric sources, the symmetric solutions exist, we assume the 2PN solution as
\begin{equation}
\begin{aligned}
h_{00}^{(4)}&=U_3(r),\\
h_{ij}^{(4)}&=U_4(r)\delta_{ij}.\label{eq:U34def}
\end{aligned}
\end{equation}
Thus the equation for $U_3$ and $U_4$ is given as
\begin{equation}
\begin{aligned}
-\frac{1}{2}\Delta U_3 
+\frac{1}{3}(2\lambda+3\mu)\Delta\Delta U_3
+\frac{2}{3}(\lambda-3\mu)\Delta\Delta U_4+S_1=0,\label{eq:O4U1}
\end{aligned}
\end{equation}
\begin{equation}
\begin{aligned}
&\partial_a\partial_b\left(
\frac{1}{2}U_3-\frac{1}{2}U_4
-\frac{1}{3}(\lambda+6\mu)\Delta U_3
+\frac{1}{3}(-\lambda+12\mu) \Delta U_4
\right)
\\&
+\delta_{ab}\left(
-\frac{1}{2}\Delta U_4
+\frac{1}{3}(\lambda-3\mu)\Delta\Delta U_3
+\frac{1}{3}(\lambda+6\mu)\Delta\Delta U_4
\right)
\\&
+A_{ab}+B\delta_{ab}=0.\label{eq:O4U2}
\end{aligned}
\end{equation}
Take the trace of \eq{eq:O4U2}, we get
\begin{equation}
\begin{aligned}
&
\frac{1}{2}\Delta U_3 -2\Delta U_4 
+\frac{1}{3}(2\lambda-15\mu)\Delta\Delta U_3
\\&
+\frac{2}{3}(\lambda+15\mu)\Delta\Delta U_4
+S_2=0,\label{eq:O4U3}
\end{aligned}
\end{equation}
where $S_2 = A_{ab}\delta_{ab}+3B$.
We use the same method used in Sec.~\ref{sec:1PN} to solve the equations~(\ref{eq:O4U1}) and~(\ref{eq:O4U3}).
The Fourier transformations of equations~(\ref{eq:O4U1}) and~(\ref{eq:O4U3}) are
\begin{equation}
\begin{aligned}
% \frac{1}{6}k^2\left(3+\frac{m_W^2+2m_R^2}{m_R^2+m_W^2}k^2\right)\tilde{U}_3
% +\frac{-m_W^2+m_R^2}{3(m_R^2+m_W^2)}k^4\tilde{U}_
&
\frac{1}{6}k^2(3+(4\lambda+6\mu)k^2)\tilde{U}_3
\\&
+\frac{2}{3}(\lambda-3\mu)k^4\tilde{U}_4+\tilde{S}_1=0,
\\&
\frac{1}{6}k^2(-3+(4\lambda-30\mu)k^2)\tilde{U}_3
\\&
+\frac{2}{3}k^2(3+(\lambda+15\mu)k^2)\tilde{U}_4
+\tilde{S}_2=0,
\end{aligned}
\end{equation}
and the solutions are
\begin{equation}
\begin{aligned}
\tilde{U}_3&=\frac{-2\tilde{S}_1}{k^2}
+\frac{\tilde{S}_1-\tilde{S}_2}{3(k^2+m_R^2)}
+\frac{5\tilde{S}_1+\tilde{S}_2}{3(k^2+m_W^2)},
\\
\tilde{U}_4&=\frac{-\tilde{S}_1-\tilde{S}_2}{2k^2}
+\frac{-\tilde{S}_1+\tilde{S}_2}{3(k^2+m_R^2)}
+\frac{5\tilde{S}_1+\tilde{S}_2}{6(k^2+m_W^2)},
\end{aligned}
\end{equation}
where $m_W^2=\frac{1}{2\lambda}$ and $m_R^2=\frac{1}{6\mu}$.
To obtain the solution for $U_3$ and $U_4$ in position space, we need to calculate the convolution between the kernel $1/r$, $e^{-m_W r}/r$, and $e^{-m_R r}/r$, and the sources $S_1$ and $S_2$.

For a static point-like source~(\ref{eq:source}), the term proportional to $\Pi$, $p$, $v_a$ and $\mathcal{V}_a$ in the sources $S_1$ and $S_2$ are set to be zero.
Also, terms with time derivatives are also zero.
% Using the solution for $U_1$ and $U_2$ as \eq{eq:solU12}, 
% and the properties \eq{eq:potential_relation}, we can reduce the term $\Delta U_1$ and $\Delta U_2$ into a linear combination of $U$, $U_R$, $U_W$ and $\rho$.
Since the terms proportional to $\rho$ are related to the gravitational self-energy and belong to the term $\Phi^{(4)}(r)$,
we also drop these terms in the calculation of $\beta(r)$.
Finally, we find that the related terms in $S_1$ and $S_2$ are proportional to the term with form $U_i U_j$, $\partial_a U_i \partial_a U_j$, and $\partial_a\partial_b U_i \partial_a\partial_b U_j$, where $U_i\in \{U, U_W, U_R\}$.
In the  point-like source, $U=M/r$, $U_W=Me^{-m_W r}/r$, and $U_R=Me^{-m_R r}/r$.
With relations %$\partial_a f(r) \partial_a g(r)$
\begin{equation}
\begin{aligned}
\partial_a f(r) \partial_a g(r) &= f'(r)g'(r),
\\
\partial_a\partial_b f(r) \partial_a\partial_b g(r) &=f''(r)g''(r)+\frac{2}{r^2}f'(r)g'(r),
\end{aligned}
\end{equation}
the expressions for $S_1$ and $S_2$ are

\begin{widetext}

\begin{equation}
\begin{aligned}
S_1=&
-\frac{\kappa^2 }{32\pi^2 }\frac{M^2}{r^4}
+\frac{\kappa^2 (m_R^2+5m_W^2)}{16\pi^2 m_R^2 m_W^2}\frac{M^2}{r^6}
+\frac{\kappa^2 M^2 e^{-m_R r}}{\pi^2} \left(
\frac{4 m_R^2-m_W^2}{12  m_R^2 m_W^2 r^6}
+\frac{4 m_R^2-m_W^2}{12  m_R m_W^2 r^5}
+\frac{2 m_R^2-m_W^2}{12  m_W^2 r^4}
\right.\\&\left.
+\frac{m_R^3-m_R m_W^2}{18  m_W^2 r^3}
+\frac{4 m_R^4-7 m_R^2 m_W^2}{288  m_W^2 r^2}
\right)
+\frac{\kappa^2 M^2 e^{- m_W r}}{\pi^2} \left(
\frac{-11 m_R^2-13 m_W^2}{24  m_R^2 m_W^2 r^6}
+\frac{-11 m_R^2-13 m_W^2}{24  m_R^2 m_W r^5}
\right.\\&\left.
+\frac{-5 m_R^2-27 m_W^2}{96  m_R^2 r^4}
+\frac{29 m_W\left(m_R^2-m_W^2\right)}{288  m_R^2 r^3}
+\frac{5 m_R^2 m_W^2-2 m_W^4}{72   m_R^2 r^2}
\right)
+\frac{\kappa^2 M^2 e^{-2m_R r}}{\pi^2} \left(
\frac{1}{48  m_R^2 r^6}
+\frac{1}{24  m_R r^5}
\right.\\&\left.
+\frac{5 m_R}{144 r^3}
+\frac{5 m_R^2}{192 r^2}
+\frac{13}{288 r^4}
\right)
+\frac{\kappa^2 M^2 e^{-2 m_W r}}{\pi^2} \left(
\frac{19 m_R^2+12 m_W^2}{48  m_R^2 m_W^2 r^6}
+\frac{19 m_R^2+12 m_W^2}{24 m_R^2 m_W r^5}
+\frac{93 m_R^2+74 m_W^2}{144 m_R^2  r^4}
\right.\\&\left.
+\frac{17 m_R^2 m_W+26 m_W^3}{72 m_R^2  r^3}
-\frac{7 m_W^2 \left(m_R^2-4 m_W^2\right)}{144 m_R^2 r^2}
\right)
+\frac{\kappa^2 M^2 e^{-(m_R+m_W) r}}{\pi^2} \left(
\frac{m_W^2-8 m_R^2}{24  m_R^2 m_W^2 r^6}
\right.\\&\left.
+\frac{-8 m_R^2 m_W+m_R m_W^2-8 m_R^3+m_W^3}{24  m_R^2 m_W^2 r^5}
+\frac{-96 m_R^3 m_W-31 m_R^2 m_W^2+12 m_R m_W^3-48 m_R^4+5 m_W^4}{288  m_R^2 m_W^2 r^4}
\right.\\&\left.
+\frac{-48 m_R^4 m_W-35 m_R^3 m_W^2+m_R^2 m_W^3+5 m_R m_W^4-16 m_R^5+m_W^5}{288  m_R^2 m_W^2 r^3}
\right.\\&\left.
-\frac{(8 m_R^2 m_W-2 m_R m_W^2+4 m_R^3-m_W^3) (m_R+m_W)^2}{288  m_R m_W^2 r^2}
\right),
\end{aligned}
\end{equation}

\begin{equation}
\begin{aligned}
S_2=&
\frac{\kappa^2 }{8\pi^2 }\frac{M^2}{r^4}
+\frac{\kappa^2 (m_R^2-25m_W^2)}{16\pi^2 m_R^2 m_W^2}\frac{M^2}{r^6}
+\frac{\kappa^2 M^2 e^{-m_R r}}{\pi^2} \left(
\frac{4 m_R^2+5 m_W^2}{12 m_R^2 m_W^2 r^6 }
+\frac{4 m_R^2+5 m_W^2}{12 m_R m_W^2 r^5 }
+\frac{2 m_R^2+3 m_W^2}{12  m_W^2 r^4}
\right.\\&\left.
+\frac{2 m_R m_W^2+m_R^3}{18 m_W^2  r^3}
+\frac{11 m_R^2 m_W^2+4 m_R^4}{288 m_W^2  r^2}
\right)
+\frac{\kappa^2 M^2 e^{-m_W r}}{\pi^2} \left(
\frac{65 m_W^2-11 m_R^2}{24 m_R^2 m_W^2 r^6}
+\frac{65 m_W^2-11 m_R^2}{24 m_R^2 m_W r^5}
\right.\\&\left.
+\frac{135 m_W^2-31 m_R^2}{96 m_R^2 r^4}
+\frac{145 m_W^3-49 m_R^2 m_W}{288 m_R^2 r^3}
+\frac{10 m_W^4-7 m_R^2 m_W^2}{72 m_R^2 r^2}
\right)
+\frac{\kappa^2 M^2 e^{-2 m_R r}}{\pi^2} \left(
-\frac{5}{48 m_R^2 r^6}
\right.\\&\left.
-\frac{5}{24 m_R r^5}
-\frac{7 m_R}{36 r^3}
-\frac{25 m_R^2}{192 r^2}
-\frac{17}{72 r^4}
\right)
+\frac{\kappa^2 M^2 e^{-2 m_W r}}{\pi^2} \left(
\frac{19 m_R^2-60 m_W^2}{48 m_R^2 m_W^2  r^6}
+\frac{19 m_R^2-60 m_W^2}{24 m_R^2 m_W r^5}
\right.\\&\left.
+\frac{37 \left(3 m_R^2-10 m_W^2\right)}{144 m_R^2 r^4}
+\frac{5 \left(7 m_R^2 m_W-26 m_W^3\right)}{72 m_R^2 r^3}
+\frac{35 m_W^2 \left(m_R^2-4 m_W^2\right)}{144 m_R^2  r^2}
\right)
\\&
+\frac{\kappa^2 M^2 e^{-(m_R+m_W) r}}{\pi^2} \left(
\frac{-8 m_R^2-5 m_W^2}{24 r^6 m_R^2 m_W^2}
+\frac{-8 m_R^2 m_W-5 m_R m_W^2-8 m_R^3-5 m_W^3}{24 r^5 m_R^2 m_W^2}
\right.\\&\left.
+\frac{-96 m_R^3 m_W-85 m_R^2m_W^2-60 m_R m_W^3-48 m_R^4-25 m_W^4}{288 r^4 m_R^2 m_W^2}
\right.\\&\left.
+\frac{-48 m_R^4 m_W-65 m_R^3 m_W^2-53 m_R^2 m_W^3-25 m_R m_W^4-16 m_R^5-5m_W^5}{288 r^3 m_R^2 m_W^2}
\right.\\&\left.
-\frac{(8 m_R^2 m_W+10 m_R m_W^2+4 m_R^3+5 m_W^3) (m_R+m_W)^2}{288 r^2 m_R m_W^2}
\right).
\end{aligned}
\end{equation}

\end{widetext}

From the expressions of $S_1$ and $S_2$, we find that we need to calculate the following convolutions:
\begin{itemize}
\item $\frac{1}{r}*\frac{1}{r^n}$, with $n=4,6$;
\item $\frac{1}{r}*\frac{e^{-m r}}{r^n}$, with $n=2,3,4,5,6$;
\item $\frac{e^{-m r}}{r}*\frac{1}{r^n}$, with $n=4,6$;
\item $\frac{e^{-m_1 r}}{r}*\frac{e^{-m_2 r}}{r^n}$, with $n=2,3,4,5,6$.
\end{itemize}
In the following, we will calculate the convolutions and get $\beta(r)$.

\subsection{Massless sector}

In this section, we calculate the convolution of $\frac{1}{r}*\frac{1}{r^n}$ and $\frac{1}{r}*\frac{e^{-m r}}{r^n}$, and calculate their contributions to $h_{00}^{(4)}$ and $h_{ab}^{(4)}$.
Unfortunately, most of the convolution integrals are divergent because of the point-like limit of the source, and the divergences originate from the gravitational self-energy.
Take the GR case as an example, in the limit $m_W\to\infty$ and $m_R\to\infty$, $S_1=-\frac{\kappa^2}{32\pi^2}\partial_a U \partial_a U=-2G^2\partial_a U \partial_a U$, and we have
\begin{equation}
\begin{aligned}
h_{00}^{(4)}&=
\frac{1}{4\pi r}*(-2S_1)=4G^2\frac{1}{4\pi r}*(\partial_a U \partial_a U)
\\&
=4G^2\frac{1}{4\pi r}*\left(\frac{1}{2}\Delta(U^2)-U\Delta U\right)
\\&
=2G^2\Delta\left(\frac{1}{4\pi r}\right)*U^2+4G^2 \frac{1}{r}*(\rho U)
\\&
=-2G^2 \delta^{(3)}(x)*U^2+4G^2 \frac{1}{r}*(\rho U)
\\&
=-2G^2U^2+4G^2 \frac{1}{r}*\frac{M^2\delta^{(3)}(x)}{r}.
\end{aligned}
\end{equation}
The first part gives $\beta=1$, and the second part is the gravitational self-energy.
In the case of $\rho=M\delta^{(3)}(x)$, it diverges.
Therefore, it is necessary to adopt a regularization scheme to subtract the infinities arising from the contributions of gravitational self-energy within these convolution integrals.

However, in this sector, we have a tricky way to simplify the regularization scheme.
Since in the limit $m_W\to\infty$ and $m_R\to\infty$, the contribution for $h_{00}^{(4)}$ in this sector should reduce to the GR case $-\frac{2G^2M^2}{r^2}$,
thus we can safely neglect any contribution of order higher than $1/r^2$ after the regularization.
On the other hand, the convolutions $u(r)=\frac{1}{4\pi r}*f(r)$ satisfies the equation $\Delta u=-f$, and thus
\begin{equation}
u''(r)+\frac{2}{r}u'(r)=-f(r),
\end{equation}
and the solution is
\begin{equation}
u(r)=-\int\frac{ dr'}{r'^2}\int^{r'} f(r'')r''^2 dr''+\frac{c_1}{r}+c_2.
\end{equation}
Since $\frac{c_1}{r}$ and $c_2$ are higher than $1/r^2$, 
% \lh{$c_0$???}
we can safely drop them.
% In other words, we adopt a regularization scheme that discards both the divergences and the $1/r$ contributions.
By employing the technique of solving convolution integrals via ordinary differential equations, we obtain the following convolutions: 
% \begin{equation}
% \begin{aligned}
% \frac{1}{4\pi r}*\frac{1}{r^4}=&-\frac{1}{2 r^2},\\
% \frac{1}{4\pi r}*\frac{1}{r^6}=&-\frac{1}{12 r^4},\\
% \frac{1}{4\pi r}*\frac{e^{-mr}}{r^2}=&-\frac{e^{-mr}}{mr}-\Ei(-mr),\\
% \frac{1}{4\pi r}*\frac{e^{-mr}}{r^3}=&\frac{e^{-mr}}{r}+(m+\frac{1}{r})\Ei(-mr),\\
% \frac{1}{4\pi r}*\frac{e^{-mr}}{r^4}=&
% -\frac{1+mr}{2r^2}e^{-mr}
% \\&
% -\frac{m(mr+2)}{2r}\Ei(-mr),\\
% \frac{1}{4\pi r}*\frac{e^{-mr}}{r^5}=&
% \frac{-1+2mr+m^2r^2}{6r^3}e^{-mr}
% \\&
% -\frac{m^2(mr+3)}{6r}\Ei(-mr),\\
% \frac{1}{4\pi r}*\frac{e^{-mr}}{r^6}=&
% -\frac{m^3 r^3+3 m^2 r^2-2 m r+2}{24 r^4}e^{-mr}
% \\&
% -\frac{m^3 (m r+4)}{24 r}\Ei(-mr),\\
% \end{aligned}
% \end{equation}
% where $\Ei(r)=-\int_{-r}^\infty \frac{e^{-t}}{t} dt$ denotes the exponential integral.
\begin{equation}
\frac{1}{4\pi r}*\frac{1}{r^n}=-\frac{1}{(n-2)(n-3)r^{n-2}},
\end{equation}
\begin{equation}
\frac{1}{4\pi r}*\frac{e^{-m r}}{r^n}=\frac{1}{r^{n-2}}\left(\E_{n-1}(mr)-\E_{n-2}(mr)\right),
\end{equation}
where $E_n(x)=\int_1^\infty \frac{e^{-x t}}{t^n}dt$ is the generalized exponential integral.
% It follows from the asymptotic behavior of the exponential integral in the case $x\gg 1$, we have
% \begin{equation}
% \mathrm{Ei}(-x)\approx\frac{e^{-x}}{x}\left(1-\frac{1!}{x}+\frac{2!}{x^2}-\frac{3!}{x^3}+\ldots\right),
% \end{equation}
It follows from the asymptotic behavior of the generalized exponential integral in the case $|x|\gg 1$, that we have
\begin{equation}
\E_n(x)=\frac{e^{-x}}{x}\sum_{k=0}^\infty(-1)^k \frac{(n)_k}{x^k},
\end{equation}
where $(n)_k$ is the Pochhammer symbol,
\begin{equation}
(n)_k=n(n+1)\cdots (n+k-1), \quad (n)_0=1.
\end{equation}
Thus, the asymptotic behavior of the convolution is
% \begin{equation}
% \frac{1}{4\pi r}*\frac{e^{-mr}}{r^k}\approx e^{-mr}\left( -\frac{1}{m^2 r^k}+\frac{2(k-1)}{m^3 r^{k+1}}+\ldots\right).
% \end{equation}
\begin{equation}
\frac{1}{4\pi r}*\frac{e^{-mr}}{r^n}=-\frac{e^{-mr}}{m^2 r^n}\sum_{k=0}^\infty (-1)^k \frac{(k+1)(n+k-2)!}{(n-2)! (mr)^k}.
\end{equation}
With the above convolution results, we can, in principle, add the convolutions together and obtain the closed form of the contributions of the massless sector.
However, in the region $m_W r\gg 1$ and $m_R r \gg 1$, it is more convenient to consider the asymptotic behavior.
In the $m_R\simeq m_W$ region, with contributions smaller than $\frac{e^{-m_W r}}{r^2}$ and  $\frac{e^{-m_R r}}{r^2}$ ignored,
the results are
\begin{equation}
\begin{aligned}
U_3^{(0)}\approx&
-\frac{\kappa ^2 M^2}{32 \pi ^2 r^2}
+\frac{\kappa ^2 M^2 \left(m_R^2+5 m_W^2\right)}{96 \pi ^2 m_R^2 m_W^2  r^4}
\\&
+\frac{\kappa ^2 M^2 \left(4 m_R^2-7 m_W^2\right)}{144 \pi^2  m_W^2} \frac{e^{- m_R r}}{r^2}
\\&
+\frac{\kappa ^2 M^2  \left(5 m_R^2-2 m_W^2\right)}{36 \pi ^2  m_R^2} \frac{e^{- m_W r}}{r^2},\\
% \end{aligned}
% \end{equation}
% \begin{equation}
% \begin{aligned}
U_4^{(0)}\approx&
\frac{3 \kappa ^2 M^2}{128 \pi ^2 r^2}
+\frac{\kappa ^2 M^2 \left(m_R^2-10 m_W^2\right)}{192 \pi ^2  m_R^2 m_W^2 r^4}
\\&
+\frac{\kappa ^2 M^2  \left(2 m_R^2+m_W^2\right)}{144 \pi ^2  m_W^2}\frac{e^{- m_R r}}{r^2}
\\&
-\frac{\kappa ^2 M^2  \left(m_R^2-4 m_W^2\right)}{72 \pi^2  m_R^2}\frac{e^{- m_W r}}{r^2}.\label{eq:U340}
\end{aligned}
\end{equation}
In the limit $m_R\to\infty$ and $m_W\to\infty$, the result reduces to the GR case.

We also consider the case that $\lambda=0$ or $\mu=0$.
The result can be obtained either by taking the appropriate limit of \eq{eq:U340}, or by first taking the limit of the source before performing the convolution. 
We have explicitly verified that the two procedures lead to identical results.
When $\lambda=0$, $m_W\to\infty$, the theory reduces to $f(R)$ theory, and the result is
\begin{equation}
\begin{aligned}
U_3^{(0)}\approx&
-\frac{\kappa ^2 M^2}{32 \pi ^2 r^2}
+\frac{5 \kappa ^2 M^2  }{96 \pi ^2 m_R^2  r^4}
\\&
-\frac{7 \kappa ^2 M^2 }{144 \pi^2  } \frac{e^{- m_R r}}{r^2},
\\
U_4^{(0)}\approx&
\frac{3 \kappa ^2 M^2}{128 \pi ^2 r^2}
-\frac{5 \kappa ^2 M^2  }{96 \pi ^2  m_R^2  r^4}
\\&
+\frac{\kappa ^2 M^2  }{144 \pi ^2 }\frac{e^{- m_R r}}{r^2}.
\end{aligned}
\end{equation}
And when $\mu=0$, $m_R\to\infty$, the result is
\begin{equation}
\begin{aligned}
U_3^{(0)}\approx&
-\frac{\kappa ^2 M^2}{32 \pi ^2 r^2}
+\frac{ \kappa ^2 M^2  }{96 \pi ^2 m_W^2  r^4}
\\&
+\frac{5 \kappa ^2 M^2 }{36 \pi^2  } \frac{ e^{- m_W r}}{r^2},
\\
U_4^{(0)}\approx&
\frac{3 \kappa ^2 M^2}{128 \pi ^2 r^2}
+\frac{ \kappa ^2 M^2  }{192 \pi ^2  m_W^2  r^4}
\\&
-\frac{\kappa ^2 M^2  }{72 \pi ^2 }\frac{e^{- m_W r}}{r^2}.
\end{aligned}
\end{equation}

\subsection{Massive sector}

In this sector, we need to adopt a regularization scheme to calculate the convolution of $\frac{e^{-m r}}{r}*\frac{1}{r^n}$ and $\frac{e^{-m_1 r}}{r}*\frac{e^{-m_2 r}}{r^n}$,
and we adopt a scheme analogous to dimensional regularization.

In three-dimensional Euclidean space, the convolution of $f(r)$ and $g(r)$ can be expressed as 
\begin{equation}
(f*g)(r)=\frac{2\pi}{r}\int_0^\infty dr^{\prime}r^{\prime}g(r^{\prime})\int_{|r-r^{\prime}|}^{r+r^{\prime}}ds\left.s\right.f(s).\label{eq:int}
\end{equation}
For the convolution $\frac{e^{-m r}}{r}*\frac{1}{r^n}$, we denote $f=\frac{1}{r^n}$ and $g=\frac{e^{-m r}}{r}$ and perform the integration~(\ref{eq:int}),
and the result is
\begin{equation}
\begin{aligned}
\frac{e^{-mr}}{4\pi r}*\frac{1}{r^n}=&
\frac{e^{mr}E_{n-1}(mr)-e^{-mr}E_{n-1}(-mr)}{2mr^{n-1}}\\
&+\left((-1)^n-1\right)\frac{m^{n-3}e^{-mr}}{2r}\Gamma(2-n),
\end{aligned}
\end{equation}
where $\Gamma(x)=\int_0^\infty t^{x-1}e^{-t}$ is the Gamma function.
The divergence appears in $\Gamma(2-n)$, since negative integers are poles of the Gamma function.
However, since we only need the results for $n=4$ and $n=6$, and when $n$ is an even integer, the prefactor $(-1)^n-1=0$, thus the divergence disappears.
So the convolution result is 
\begin{equation}
\frac{e^{-mr}}{4\pi r}*\frac{1}{r^n}=
\frac{e^{mr}\E_{n-1}(mr)-e^{-mr}\E_{n-1}(-mr)}{2mr^{n-1}},\label{eq:k2rn}
\end{equation}
and the asymptotic behavior of the convolution is
\begin{equation}
\frac{e^{-mr}}{4\pi r}*\frac{1}{r^n}=\frac{1}{m^2r^n}\sum_{k=0}^\infty\frac{(n-1)_{2k}}{(mr)^{2k}}.
\end{equation}
Surprisingly, the contribution from this convolution is of polynomial order.

For the convolution $\frac{e^{-m_1 r}}{r}*\frac{e^{-m_2 r}}{r^n}$,
we denote $f=\frac{e^{-m_2 r}}{r^n}$ and $g=\frac{e^{-m_1 r}}{r}$ and perform the integration~(\ref{eq:int}),
and the result is
\begin{equation}
\begin{aligned}
&\frac{e^{-m_1 r}}{4\pi r}*\frac{e^{-m_2 r}}{r^n}\\
=&
\frac{e^{m_1 r}\E_{n-1}((m_2+m_1)r)-e^{m_1 r}\E_{n-1}((m_2-m_1)r)}{2m_1 r^{n-1}}\\
% -\frac{e^{m_1 r}\E_{n-1}((m_2-m_1)r)}{2m_1 r^{n-1}}\\
&+\frac{(m_2-m_1)^{n-2}-(m_2+m_1)^{n-2}}{2m_1 r}\Gamma(2-n)e^{-m_1 r}.\label{eq:k2m1m2}
\end{aligned}
\end{equation}
Not surprisingly, when $m_2=0$ and $m_1=m$, the result reduces to \eq{eq:k2rn}. 
Still, the divergence appears in $\Gamma(2-n)$.
To subtract the divergence, we use the fact that 
\begin{equation}
\Gamma(-n+\epsilon)=\frac{(-1)^n}{n!}\left(\frac{1}{\epsilon}+\psi(n+1)\right)+\mathcal{O}(\epsilon),
\end{equation}
where
$\psi(x)=\frac{d}{dx}\ln\Gamma(x)$ is the digamma function, and
\begin{equation}
\psi(n+1)=-\gamma_{\rm E}+H_n,\quad H_n=\sum_{k=1}^n\frac{1}{k},
\end{equation}
where $\gamma_{\rm E} = 0.57721\ldots$ is the Euler gamma constant.
In principle, we can use the minimal subtraction scheme to remove the divergent part $\frac{1}{\epsilon}$.
However, here we keep the $\frac{1}{\epsilon}$ part in our calculation to trace the behavior of the divergence, that is, we use
\begin{equation}
\Gamma(-n+\epsilon)=\frac{(-1)^{n}}{n!}\left(\frac{1}{\epsilon}-\gamma_{\rm E}+H_{n}\right).\label{eq:scheme}
\end{equation}
The asymptotic behavior of this convolution is 
\begin{equation}
\begin{aligned}
&\frac{e^{-m_1 r}}{4\pi r}*\frac{e^{-m_2 r}}{r^k}\\
\approx& \frac{(m_2-m_1)^{n-2}-(m_2+m_1)^{n-2}}{2m_1 }\Gamma(2-n)\frac{e^{-m_1 r}}{r}\\
&+\frac{e^{-m_2 r}}{r^n}\left(\frac{1}{m_1^2-m_2^2}+\frac{2(n-1)m_2}{(m_1^2+m_2^2)^2}\frac{1}{r}+\ldots\right).
\end{aligned}
\end{equation}

For the convolution $\frac{e^{-m r}}{r}*\frac{e^{-m r}}{r^n}$,
we adopt the limit $m_1\to m$ and $m_2 \to m$ of \eq{eq:k2m1m2}.
Using the fact that $E_n(0)=1/(n-1)$ for $n>1$, we have
\begin{equation}
\begin{aligned}
\frac{e^{-mr}}{4\pi r}*\frac{e^{-mr}}{r^n}=&
\frac{e^{m r}\E_{n-1}(2mr)}{2m r^{n-1}}
-\frac{e^{-mr}}{2(n-2)m r^{n-1}}\\
&-(2m)^{n-3}\Gamma(2-n)\frac{e^{-mr}}{r}.\label{eq:k2m}
\end{aligned}
\end{equation}
The asymptotic behavior of this convolution is
\begin{equation}
\begin{aligned}
&\frac{e^{-mr}}{4\pi r}*\frac{e^{-mr}}{r^n}\\
\approx&-(2m)^{n-3}\Gamma(2-n)\frac{e^{-mr}}{r}\\
&+\frac{e^{-mr}}{2m r^{n-1}}\left(-\frac{1}{n-2} +\frac{1}{2m r}-\frac{n-1}{4m^2 r^2}+\ldots\right ).
\end{aligned}
\end{equation}
For $n=2$, we take the $n\to 2$ limit of \eq{eq:k2m}, and we get
\begin{equation}
\begin{aligned}
\frac{e^{-mr}}{4\pi r}*\frac{e^{-mr}}{r^2}=&\frac{e^{-mr}}{2mr}(\gamma_{\rm E}+\ln(2mr))+\frac{e^{mr}}{2mr}\E_1(2mr).\label{eq:k2mr2}
\end{aligned}
\end{equation}
The asymptotic behavior of this convolution is
\begin{equation}
\begin{aligned}
\frac{e^{-mr}}{4\pi r}*\frac{e^{-mr}}{r^2}=&\frac{e^{-mr}}{mr}(\gamma_{\rm E}+\ln(2mr))\\
&+\frac{e^{-mr}}{(2mr)^2}\sum_{k=0}^\infty \frac{(-1)^k k!}{(2mr)^k}.
\end{aligned}
\end{equation}

Using the convolutions~(\ref{eq:k2rn}),~(\ref{eq:k2m1m2}),~(\ref{eq:k2m}), and~(\ref{eq:k2mr2}), and the regularization scheme~(\ref{eq:scheme}), we can obtain the results from the massive sector.
Although we can obtain an exact result, it is extremely lengthy and involves a large number of exponential integrals.
Instead, we consider the asymptotic result in the region where $m_W r\gg 1$ and $m_R r \gg 1$.
It follows that most of the polynomial contributions cancel, leaving only terms of order $\mathcal{O}(1/r^4)$.
For the contributions with the exponential, we can see the leading contribution is of order $\mathcal{O}(e^{-mr}\ln r/r)$, which is from the convolution $\frac{e^{-mr}}{ r}*\frac{e^{-mr}}{r^2}$.
Although we keep the $\frac{1}{\epsilon}$ in the regularization scheme, we find that all the $\frac{1}{\epsilon}$ divergences cancel among themselves.
 % in a manner analogous to loop calculations in quantum field theory. \lh{In qft we need to include counter terms, loops cannot cancel the divergences.}
% \lh{Notably the process of cancellation of the divergences does not include any artificial choice of regularization schemes, therefore the result is natural and unique.}
Notably, the process of cancellation of the divergences does not include any artificial choice of regularization schemes; therefore the result is natural and unique.
We compute the asymptotic expansion up to $\mathcal{O}(e^{-mr}/r)$.
We first consider the region $m_R\simeq m_W$. For the $R^2$ sector, the result is
\begin{equation}
\begin{aligned}
U_3^{(R)}=&\frac{e^{-m_R r}}{4\pi r}*\frac{S_1-S_2}{3}\\
\approx&
-\frac{5\kappa^2M^2}{96\pi^2 m_R^2 r^4}-\frac{\kappa^2M^2m_R}{96\pi^2}\frac{e^{-m_R r}\ln(2 m_R r)}{r}\\
&-\frac{\kappa^2 M^2e^{-m_R r}}{288 \pi^2 m_R^2 r}(3(1+\gamma_{\rm E})m_R^3+5m_R^2 m_W +16m_W^3),\\
U_4^{(R)}=&\frac{e^{-m_R r}}{4\pi r}*\frac{-S_1+S_2}{3}\\
=&-U_3^{(R)}.\label{eq:U34R}
\end{aligned}
\end{equation}
For the Weyl sector, the result is
\begin{equation}
\begin{aligned}
U_3^{(W)}=&\frac{e^{-m_W r}}{4\pi r}*\frac{5S_1+S_2}{3}\\
\approx&
-\frac{\kappa^2M^2}{96\pi^2 m_W^2 r^4}+\frac{\kappa^2M^2m_W}{24\pi^2}\frac{e^{-m_W r}\ln(2 m_W r)}{r}\\
&+\frac{\kappa^2 M^2e^{-m_W r}}{288\pi^2 m_W r}(8 m_R^2+9 m_R m_W +(7+12 \gamma_{\rm E})m_W^2),\\
U_4^{(W)}=&\frac{e^{-m_W r}}{4\pi r}*\frac{5S_1+S_2}{6}\\
=&\frac{1}{2}U_3^{(W)}.\label{eq:U34W}
\end{aligned}
\end{equation}

We also consider the case that $\lambda=0$ or $\mu=0$.
In this case, taking the limit directly on the above expression is not well defined. We therefore choose to first take the limit of the source and then perform the convolution.
When $\lambda=0$, $m_W\to\infty$, we only have contributions from the $R^2$ sector, and the result is
\begin{equation}
\begin{aligned}
U_3^{(R)}\approx&-\frac{5\kappa^2M^2}{96\pi^2 m_R^2 r^4}
-\frac{(9\gamma_{\rm E}-4)\kappa^2 M^2 m_R e^{-m_R r}}{864 \pi^2r}\\
&-\frac{\kappa^2M^2m_R}{96\pi^2}\frac{e^{-m_R r}\ln(2 m_R r)}{r},\\
U_4^{(R)}=&-U_3^{(R)}.
\end{aligned}
\end{equation}
When $\mu=0$, $m_R\to\infty$, we only have contributions from the Weyl sector, and the result is
\begin{equation}
\begin{aligned}
U_3^{(W)}
\approx&
-\frac{\kappa^2M^2}{96\pi^2 m_R^2 r^4}
+\frac{(13+36\gamma_{\rm E})\kappa^2 M^2 m_W e^{-m_W r}}{864\pi^2  r}\\
&+\frac{\kappa^2M^2m_W}{24\pi^2}\frac{e^{-m_W r}\ln(2 m_W r)}{r},\\
U_4^{(W)}=&\frac{1}{2}U_3^{(W)}.
\end{aligned}
\end{equation}
Once again, the $\frac{1}{\epsilon}$ divergences cancel among themselves.

\subsection{Result}

Using the contributions from the massless sector~(\ref{eq:U340}), the $R^2$ sector~(\ref{eq:U34R}), and the Weyl sector~(\ref{eq:U34W}), we finally have the solutions for $U_3$ and $U_4$.
In the region $m_R\simeq m_W$, they are
\begin{equation}
\begin{aligned}
U_3=&U_3^{(0)}+U_3^{(R)}+U_3^{(W)}\\
\approx&
-\frac{\kappa^2 M^2}{32\pi^2 r^2}
-\frac{\kappa^2M^2m_R}{96\pi^2}\frac{e^{-m_R r}\ln(2 m_R r)}{r}\\
&+\frac{\kappa^2M^2m_W}{24\pi^2}\frac{e^{-m_W r}\ln(2 m_W r)}{r}\\
&-\frac{\kappa^2 M^2e^{-m_R r}}{288\pi^2 m_R^2 r}(3(1+\gamma_{\rm E})m_R^3+5m_R^2 m_W +16m_W^3)\\
&+\frac{\kappa^2 M^2e^{-m_W r}}{288\pi^2 m_W r}(8 m_R^2+9 m_R m_W +(7+12 \gamma_{\rm E})m_W^2),\\
\end{aligned}
\end{equation}
\begin{equation}
\begin{aligned}
U_4=&U_4^{(0)}+U_4^{(R)}+U_4^{(W)}\\
\approx&
\frac{3\kappa^2 M^2}{128\pi^2 r^2}
+\frac{\kappa^2M^2m_R}{96\pi^2}\frac{e^{-m_R r}\ln(2 m_R r)}{r}\\
&+\frac{\kappa^2M^2m_W}{48\pi^2}\frac{e^{-m_W r}\ln(2 m_W r)}{r}\\
&+\frac{\kappa^2 M^2e^{-m_R r}}{288\pi^2 m_R^2 r}(3(1+\gamma_{\rm E})m_R^3+5m_R^2 m_W +16m_W^3)\\
&+\frac{\kappa^2 M^2e^{-m_W r}}{576\pi^2 m_W r}(8 m_R^2+9 m_R m_W +(7+12 \gamma_{\rm E})m_W^2),\\
\end{aligned}
\end{equation}
and the 2PN solution for the static point-like source is
\begin{equation}
\begin{aligned}
h_{00}^{(4)}&=U_3(r),\\
h_{ij}^{(4)}&=U_4(r)\delta_{ij}.
\end{aligned}
\end{equation}
In the limit $m_R\to\infty$ and $m_W\to\infty$, with $\kappa=8\pi G$, we can see that the solution reduces to $h_{00}^{(4)}=-\frac{2G^2M^2}{r^2}$, $h_{ij}=\frac{3G^2M^2}{2r^2}\delta_{ij}$, corresponding to the PN expansion of the Schwarzschild metric in the isotropic coordinates
\begin{equation}
ds^2=-\left(\frac{1-\frac{GM}{2r}}{1+\frac{GM}{2r}}\right)^2dt^2+\left(1+\frac{GM}{2r}\right)^4\delta_{ij}dx^idx^j.
\end{equation}
One point is that though $\mathcal{O}(1/r^4)$ contributions appear in the massless sector and massive sector, in the final result, they are canceled, leaving only the contributions suppressed by the exponential.
In other words, in the weak-field approximation, up to 2PN, all deviations from GR in quadratic gravity are exponentially suppressed.
Another noteworthy point is that, in the 1PN result, the deviation from GR is of order $\mathcal{O}(e^{-mr}/r)$, while in the 2PN result, the leading deviation is of order $\mathcal{O}(e^{-mr}\ln r/r)$, higher than the 1PN result.
Finally, with the relation $-2 G_{\rm eff}^2\beta (r) U_N^2=U_3$ and $\kappa=8\pi G$, in the region $m_R\simeq m_W$, we get
\begin{equation}
\begin{aligned}
G_{\rm eff}^2\beta (r)
\approx& 1
+\frac{1}{3}m_R r e^{-m_R r}\ln(2 m_R r)
\\&
-\frac{4}{3}m_W r e^{-m_W r}\ln(2 m_W r)
\\&
+C_R m_R r e^{-m_R r}
+C_W m_W r e^{-m_W r},\label{eq:betaO2}
\end{aligned}
\end{equation}
where
\begin{equation}
\begin{aligned}
C_R=& \frac{1+\gamma_{\rm E}}{3}+\frac{5 m_W}{9m_R}+\frac{16 m_W^3}{9 m_R^3},\\
C_W=& -\frac{7+12\gamma_{\rm E}}{9}-\frac{m_R}{m_W}-\frac{8 m_R^2}{9 m_W^2}.
\end{aligned}
\end{equation}
We can, of course, write down the exact expression for $\beta(r)$ without using the asymptotic approximation, and the result is
\begin{widetext}
\begin{equation}
\begin{aligned}
G_{\rm eff}^2\beta (r)=&1
+\frac{1}{3}  m_R r e^{- m_R r} \ln \left(2  m_R r\right)
-\frac{4}{3}  m_W r e^{- m_W r} \ln \left(2  m_W r\right)
+C_R m_R r e^{-m_R r}
+C_W m_W r e^{-m_W r}
\\&
+\frac{2}{3} e^{- m_R r}
-\frac{8}{3} e^{- m_W r}
-\frac{1}{6}  m_R r e^{-2  m_R r}
+\frac{2}{3}  m_W r e^{-2  m_W r}
+\frac{1}{18} e^{-2  m_R r}
+\frac{31}{18} e^{-2  m_W r}
\\&
-\frac{7}{9} e^{-(m_R+m_W) r}
-\frac{1}{3} r^2 m_R^2 \Ei\left(-2  m_R r\right)
+\frac{4}{3} r^2 m_W^2 \Ei\left(-2  m_W r\right)
\\&
+m_W r e^{ m_W r} \left(-\frac{m_W^2-4 m_R^2 }{36 m_W^2} \Ei\left(-(2 m_R+m_W) r\right)-\frac{23}{12}  \Ei\left(-3  m_W r\right)+\frac{4}{3}  \Ei\left(-2  m_W r\right)\right)
% +r e^{ m_W r} \left(\left(\frac{m_R^2}{9 m_W}-\frac{m_W}{36}\right) \Ei\left(-(2 m_R+m_W) r\right)-\frac{23}{12} m_W \Ei\left(-3  m_W r\right)+\frac{4}{3} m_W \Ei\left(-2  m_W r\right)\right)
\\&
+m_W r e^{- m_W r} \left(\frac{m_W^2-4 m_R^2 }{36 m_W^2}\Ei\left( \left(m_W-2 m_R\right)r\right)+\frac{23}{12}  \Ei\left(- m_W r\right)\right)
% +r e^{- m_W r} \left(\frac{\left(m_W^2-4 m_R^2\right) \Ei\left(r \left(m_W-2 m_R\right)\right)}{36 m_W}+\frac{23}{12} m_W \Ei\left(- m_W r\right)+\frac{1}{9} \left(-\frac{8 m_R^2}{m_W}-9 m_R-(7+12 \gamma_{\rm E} ) m_W\right)\right)
\\&
+m_R r e^{ m_R r} \left(-\frac{m_W^2-4 m_R^2 }{18 m_R^2} \Ei\left(-(2 m_R+m_W) r\right)+\frac{1}{6}  \Ei\left(-3  m_R r\right)-\frac{1}{3}  \Ei\left(-2  m_R r\right)\right)
% +r e^{ m_R r} \left(\left(\frac{2 m_R}{9}-\frac{m_W^2}{18 m_R}\right) \Ei\left(-(2 m_R+m_W) r\right)+\frac{1}{6} m_R \Ei\left(-3  m_R r\right)-\frac{1}{3} m_R \Ei\left(-2  m_R r\right)\right)
\\&
+m_R r e^{- m_R r} \left(\frac{m_W^2-4 m_R^2 }{18 m_R^2}\Ei\left(- m_W r\right)-\frac{1}{6}  \Ei\left(- m_R r\right)\right),
% +r e^{- m_R r} \left(\frac{\left(m_W^2-4 m_R^2\right) \Ei\left(- m_W r\right)}{18 m_R}-\frac{1}{6} m_R \Ei\left(- m_R r\right)\right),
\label{eq:betafull}
\end{aligned}
\end{equation}
\end{widetext}
where $\Ei(x)$ denotes the exponential integral with definition and asymptotic expansion as
\begin{equation}
\Ei(-x)=-\int_x^\infty\frac{e^{-t}}{t}dt=\frac{e^{-x}}{x}\sum_{k=0}^\infty (-1)^k \frac{k!}{x^k},
\end{equation}
and the relations
\begin{equation}
\begin{aligned}
\E_{n+1}(x)&=\frac{1}{n}\left(e^{-x}-x\E_{n}(x)\right),\\
\E_{1}(x)&=-\Ei(-x),
\end{aligned}
\end{equation}
are used during the simplification.

We also consider the case that $\lambda=0$ or $\mu=0$.
Different from the 1PN behavior, the limits $m_W \to \infty$ and $m_R \to \infty$ cannot be straightforwardly taken at the level of the final results, and we need to take the limit at the beginning.
This phenomenon underscores the fact that the small-scale expansions for $\lambda$ and $\mu$ are asymptotic expansions, which is highly analogous to the perturbative loop expansion found in quantum field theory.
When $\lambda=0$, $m_W\to\infty$, we have
\begin{equation}
\begin{aligned}
G_{\rm eff}^2\beta (r)
\approx& 1+\frac{1}{3}m_R r e^{-m_R r}\ln(2 m_R r)\\
&+ \frac{1}{27}(9\gamma_{\rm E}-4)m_R r e^{-m_R r}.
\end{aligned}
\end{equation}
When $\mu=0$, $m_R\to\infty$, we have
\begin{equation}
\begin{aligned}
G_{\rm eff}^2\beta (r)
\approx& 1-\frac{4}{3}m_W r e^{-m_W r}\ln(2 m_W r)\\
&- \frac{1}{27}(36\gamma_{\rm E}+13)m_W r e^{-m_W r}.
\end{aligned}
\end{equation}
Together with \eq{eq:Gga}, we obtain the effective $\beta(r)$ and $\gamma(r)$ for the quadratic gravity.
The $\mathcal{O}(r \ln (r)e^{-mr})$ leading dependence of the correction to $\beta(r)$ is not surprising; such dependence has been previously manifested in scalar-tensor theories with a potential~\cite{Hohmann:2013rba, Hohmann:2017qje} and is expected to be a generic feature of theories with massive modes.

\section{Constraints from Observations}\label{sec:exp}

\add{\subsection{Constraints from the Solar System}}
In the preceding sections, we have derived expressions for the PPN parameters $\gamma(r)$ and $\beta(r)$ for a static point-like source.
In this section, we discuss the constraints from the experiments in the solar system.
Experimental tests of quadratic gravity within the Solar System have already been preliminarily explored in Ref.~\cite{Giacchini:2016nta}.
However, the analysis was restricted to effects at the 1PN order.
The PPN parameters have been measured by various high-precision experiments in the solar system~\cite{Will:2005va}.
In this work, we restrict ourselves to those measurements which provide the strictest bounds on the parameters $\gamma$ and $\beta$ with a characteristic interaction distance $r_0$.
Also, we choose the experiment with $r_0$ as small as possible.
In particular, we will use the bounds obtained from the following experiments:
\begin{description}
\item[Experiment~1] The most precise value for $\gamma$ has been obtained from the time delay of radar signals sent between Earth and the Cassini spacecraft on its way to Saturn~\cite{Bertotti:2003rm}.
The experiment suggests that $\gamma-1=(2.1\pm2.3)\times 10^{-5}.$
The radio signals were passing by the Sun at a distance of 1.6~solar radii or $r_0\approx 7.44\times 10^{-3}~\mathrm{AU}$.

\item[Experiment~2] The most well-known test of the parameter $\beta$ is the perihelion precession of Mercury~\cite{Will:2005va}, and the bound is $|2\gamma-\beta-1|<3\times10^{-3}$.
The distance is chosen as the semi-major axis of Mercury, which is $r_0\approx0.387~\mathrm{AU}$.

\item[Experiment~3] The tightest bounds on $\beta$ are obtained from lunar laser ranging experiments searching for the Nordtvedt effect, which would cause a different acceleration of the Earth and the Moon in the solar gravitational field~\cite{2010A&A...522L...5H}.
For fully conservative theories with no preferred frame effects, the bound is $4\beta-\gamma-3=(0.6\pm5.2)\times10^{-4}$.
However, the interaction distance is $r_0=1~\mathrm{AU}$ and not so small, since the effect is measured using the solar gravitational field.
\end{description}

\begin{figure}[htb]
	\centering
	\includegraphics[width=0.9\linewidth]{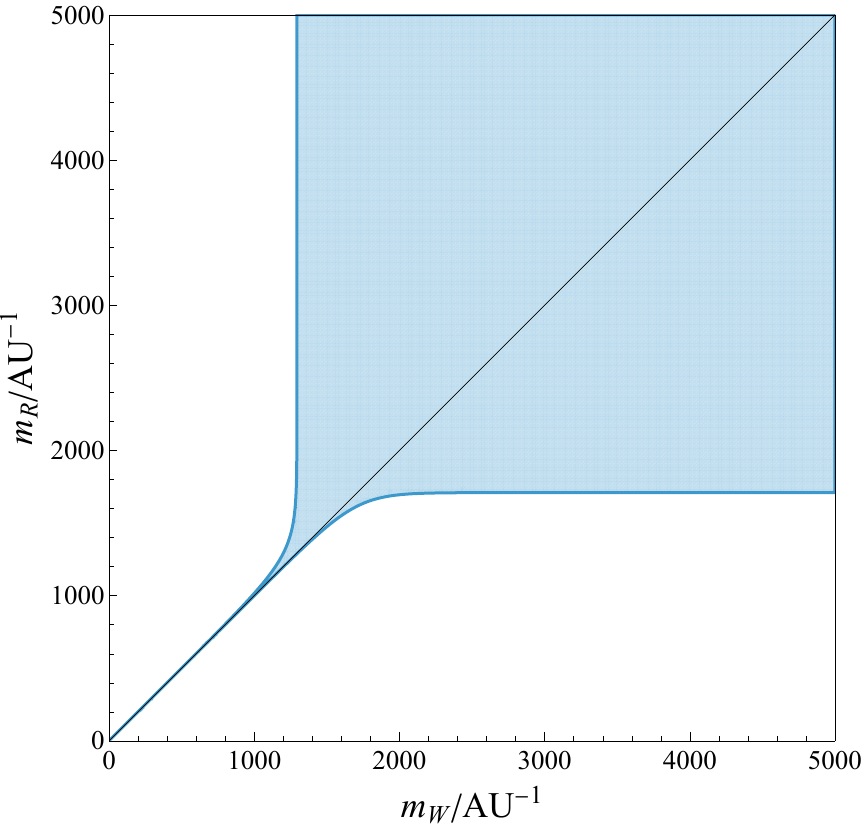}
	\caption{Allowed parameter space in the $(m_W, m_R)$ plane constrained by Experiment~1. The black line represents $m_R=m_W$.}
	\label{fig:ratio}
\end{figure}

From the expression of $\gamma(r)$ from \eq{eq:Gga}, we know that if $m_R=m_W$, then $\gamma(r)\equiv1$, which means that we cannot have a good constraint just from the constraint on the PPN parameter $\gamma$.
% Fig.~\ref{fig:ratio} shows that 
The constraint from Experiment~1 is shown in Fig.~\ref{fig:ratio}.
From Fig.~\ref{fig:ratio} we can see that if $m_W$ or $m_R$ is less than $\sim 1000~\mathrm{AU}^{-1}$, then the ratio $m_W/m_R$ should be very close to 1.
We can also obtain the constraint for the case $\lambda=0$ or $\mu =0$ from Experiment~1.
If $\lambda=0$ or $m_W\to\infty$, the constraint is given as $m_R\gtrsim 1700~\mathrm{AU}^{-1}\simeq 1.1\times10^{-8}~\mathrm{m}^{-1}$.
If $\mu=0$ or $m_R\to\infty$, the constraint is given as $m_W\gtrsim 1200~\mathrm{AU}^{-1}\simeq 8.0\times10^{-9}~\mathrm{m}^{-1}$.

\begin{figure}[htb]
	\centering
	\includegraphics[width=0.9\linewidth]{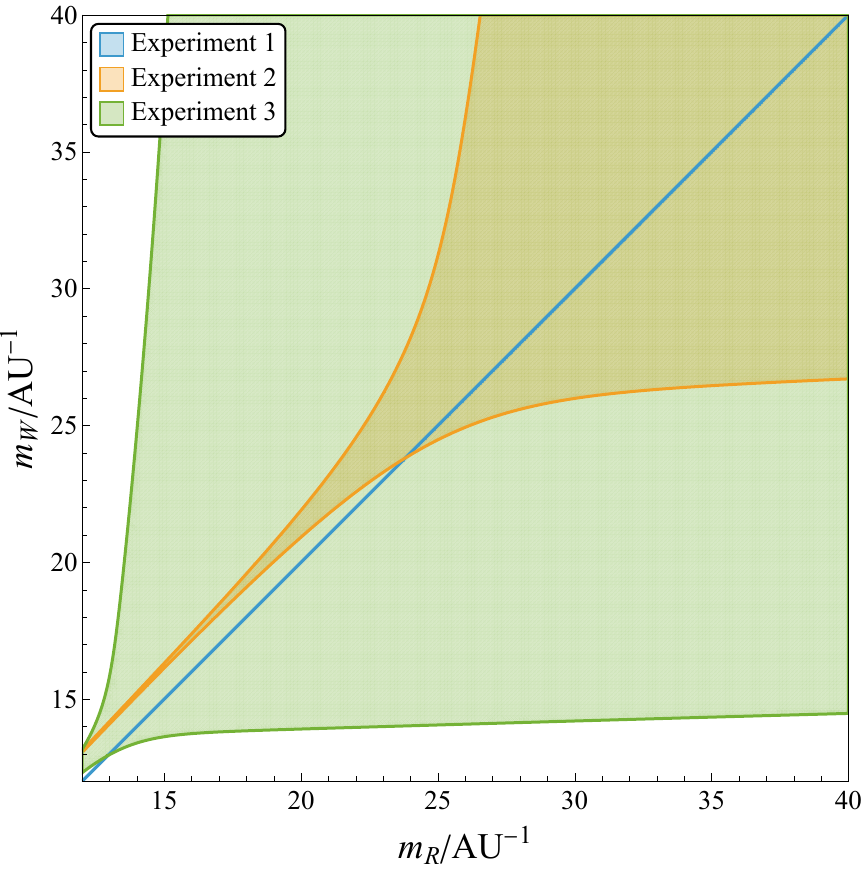}
	\caption{Allowed parameter space in the $(m_W, m_R)$ plane constrained by all of the experiments. Here, we employ the second-order approximation for $\beta(r)$ from \eq{eq:betaO2}. The blue ``line'' is the region constrained by Experiment~1 shown in Fig.~\ref{fig:ratio}.}
	\label{fig:combined}
\end{figure}

\begin{figure}[htb]
	\centering
	\includegraphics[width=0.9\linewidth]{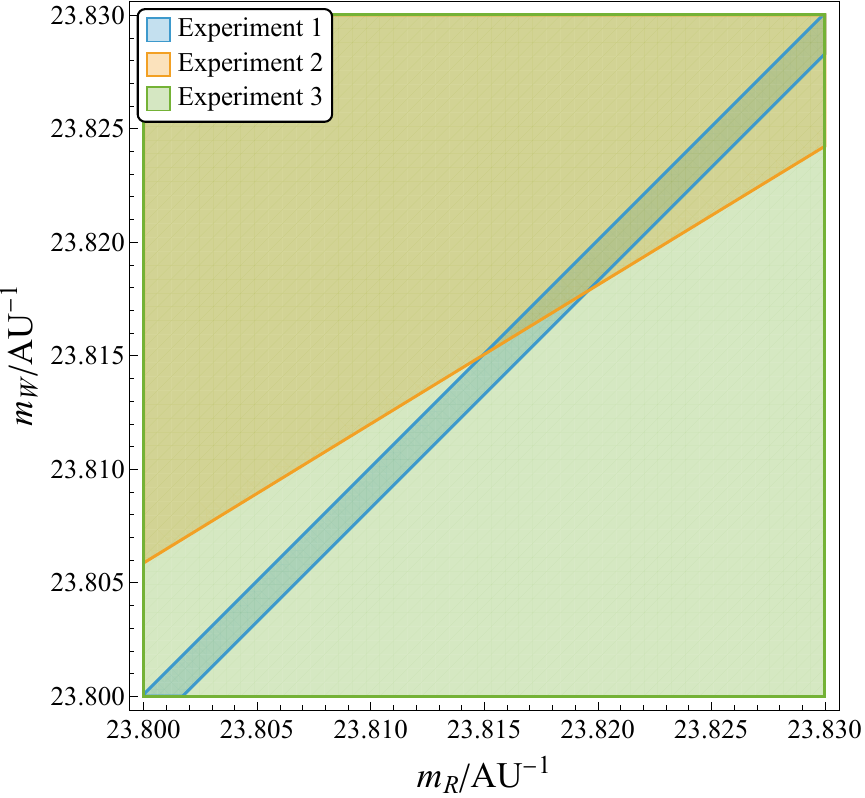}
	\caption{A zoomed-in view of Fig.~\ref{fig:combined}}
	\label{fig:zoomed}
\end{figure}

A joint analysis of the three experiments provides a preliminary constraint on the parameter space.
Using the second-order approximation of $\beta(r)$ as \eq{eq:betaO2}, the allowed parameter space in the $(m_W, m_R)$ plane constrained by the above three experiments is shown in Fig.~\ref{fig:combined}, and the overlap of these regions defines the parameter space allowed by the experiments.
Despite Experiment 2 providing a less restrictive bound on $\beta$ compared to Experiment 3, it actually imposes tighter constraints on the parameter space due to its shorter characteristic scale.
The region constrained by Experiment 1 is so narrow that it appears as a ``line'', so we provide a zoomed-in view of Fig.~\ref{fig:combined} in Fig.~\ref{fig:zoomed}.
The overlap of these regions indicates that solar-system experiments impose a constraint of $m\gtrsim 23.8~\mathrm{AU}^{-1}$, where $m$ stands for $m_R$ and $m_W$.

\begin{figure}[htb]
	\centering
	\includegraphics[width=0.9\linewidth]{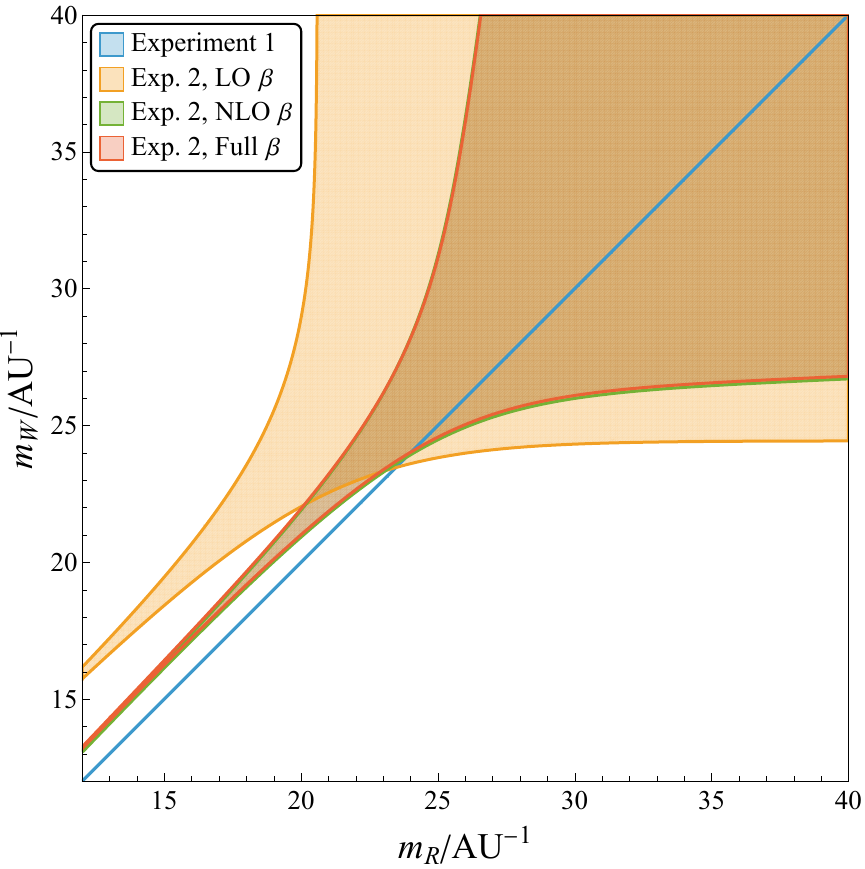}
	\caption{Allowed parameter space in the $(m_W, m_R)$ plane constrained by Experiment 1 and 2, where the first-order approximation, the second-order approximation, and the full expression of $\beta(r)$ are used, respectively. The region constrained by the second-order approximation of $\beta(r)$ almost perfectly overlaps with that of the full expression.}
	\label{fig:compare}
\end{figure}

% As a check of the validity of the asymptotic expansion of 
% $\beta(r)$, we also attempt to impose experimental constraints using only the leading-order term of $\beta(r)$ as
% \begin{equation}
% \begin{aligned}
% G_{\rm eff}^2\beta (r)
% \approx& 1
% +\frac{1}{3}m_R r e^{-m_R r}\ln(2 m_R r)
% \\&
% -\frac{4}{3}m_W r e^{-m_W r}\ln(2 m_W r),
% \end{aligned}
% \end{equation},
% and the full expression as \eq{eq:betafull}.
% and the result is shown Fig.~\ref{fig:compare}.
As a check of the validity of the asymptotic expansion of $\beta(r)$, we also attempt to impose experimental constraints using the first-order approximation, the second-order approximation, and the full expression of $\beta(r)$, respectively.
The first-order approximation of $\beta(r)$ is
\begin{equation}
\begin{aligned}
% G_{\rm eff}^2
\beta (r)
\approx& 1
+\frac{1}{3}m_R r e^{-m_R r}\ln(2 m_R r)
\\&
-\frac{4}{3}m_W r e^{-m_W r}\ln(2 m_W r),
\end{aligned}
\end{equation}
and the full expression of $\beta(r)$ is \eq{eq:betafull}.
Since the constraints from Experiment 2 are more stringent than those from Experiment 3, we have chosen the joint constraint from Experiments 1 and 2.
% Fig.~\ref{fig:compare} imposes a constraint of $m_R\gtrsim 23.2~\mathrm{AU}^{-1}$ and $m_W\gtrsim 23.2~\mathrm{AU}^{-1}$, which is close to the constraint obtained with the first two orders of $\beta(r)$, showing that using only the leading-order term of $\beta(r)$ for the constraint is sufficient.
% As a summary of the constraint, for both modes, we have $m\gtrsim 23~\mathrm{AU}^{-1}\simeq 1.5\times 10^{-10}~\mathrm{m}^{-1}$, where $m$ stands for $m_R$ or $m_W$.
Fig.~\ref{fig:compare} shows the result.
We can directly see that the region constrained by the second-order approximation of $\beta(r)$ almost perfectly overlaps with that of the full expression, which indicates that employing the second-order results is sufficient for the purpose of imposing constraints.
% In particular, considering the specific values, the constraint from the first-order approximation is $m\gtrsim 23.2~\mathrm{AU}^{-1}$, the constraint from the second-order approximation is $m\gtrsim 23.8~\mathrm{AU}^{-1}$, and the constraint from the full result is $m\gtrsim 24.0~\mathrm{AU}^{-1}$, where $m$ stands for $m_R$ and $m_W$.
Quantitatively, the first-order approximation provides a constraint of $m \gtrsim 23.4~\mathrm{AU}^{-1}$. This is refined to $23.8~\mathrm{AU}^{-1}$ at second order, which is in close agreement with the full result of $m \gtrsim 24.0~\mathrm{AU}^{-1}$.
In other words, the first-order approximation is even enough in the constraint.
Here we use the result $m \gtrsim 23~\mathrm{AU}^{-1}\simeq 1.5\times 10^{-10}{~\rm m}$,
and correspondingly, the characteristic scales of both modes are smaller than $0.043~\mathrm{AU}\simeq 6.4\times 10^9 \mathrm{m}$.
In terms of the parameters $\lambda$ and $\mu$, we have
\begin{equation}
\begin{aligned}
\lambda &=\frac{1}{2 m_W^2}&\lesssim 2.1\times 10^{19}~\mathrm{m}^2,\\
\mu&=\frac{1}{6 m_R^2}&\lesssim 7.1\times 10^{18}~\mathrm{m}^2.
\end{aligned}
\end{equation}


% \begin{figure}[htb]
% 	\centering
% 	\includegraphics[width=\linewidth]{figure/fig2.pdf}
% 	\caption{111}
% 	\label{fig1}
% \end{figure}

\add{\subsection{Comparing with GW bounds}}

\add{
Gravitational wave (GW) experiments provide an alternative approach to testing GR~\cite{Berti:2015itd, Barack:2018yly, Berti:2018cxi} and can also be used to constrain quadratic gravity.
The presence of two additional massive modes leads to extra energy loss during the inspiral phase, which in turn introduces corrections to the GW waveforms~\cite{Yagi:2011xp, Alves:2022yea, Alves:2025qcx}.
For a binary system, a meticulous analysis of the additional modes reveals that the two extra modes are excited only when $\omega > mc / 2$, where $\omega$ denotes the orbital frequency of the binary, and $m$ represents the masses of the different massive modes corresponding to $m_W$ and $m_R$~\cite{Alves:2022yea, Alves:2025qcx}.
For a binary system with a total mass $M$, the physical constraint that the orbital separation must exceed the sum of the Schwarzschild radii of the two components implies a maximum orbital frequency of $\omega \sim c^3 / (GM) \sim c / R_s$, where $R_s$ denotes the Schwarzschild radius of the compact body.
Consequently, the detection of extra modes in gravitational waves is anticipated to yield an optimal sensitivity of approximately $m \sim R_s^{-1}$.
For instance, considering the coalescence of a binary black hole system with individual masses of $2 M_\odot$, the Schwarzschild radius is approximately $6 \text{ km}$. Consequently, the most optimistic upper bound on the mass parameter is estimated to be $m \gtrsim 10^{-4} \text{ m}^{-1}$.
}

\add{
However, in the context of gravitational radiation, the massive tensor modes introduced by the term $C^2$ lead to a negative radiated power~\cite{Alves:2022yea, Alves:2024gwi, Alves:2025qcx}.
This originates from the Ostrogradsky instability associated with the $C^2$ term, which leads to a non-positive-definite energy.
In general quadratic gravity, the total radiated power suffers from a degeneracy: the massive scalar modes enhance the energy loss, while the massive tensor modes provide a negative energy flux that counteracts the former.
Consequently, when utilizing gravitational waves to constrain quadratic gravity, one inevitably encounters the problem of parameter degeneracy.
In a recent study, the authors separately considered two scenarios—GR plus a massive scalar mode and GR plus a massive tensor mode—yielding the constraints $m_R \gtrsim 2\times10^{-6}~{\rm m}^{-1}$ and $m_W \gtrsim 7\times10^{-6}~{\rm m}^{-1}$, respectively~\cite{Alves:2025qcx}.
These results are more stringent than the constraints obtained using the PPN formalism to independently limit these two scenarios ($m_R\gtrsim 1.1\times10^{-8}~\mathrm{m}^{-1}$ and $m_W\gtrsim 8.0\times10^{-9}~\mathrm{m}^{-1}$).
Despite the superior sensitivity of GW-based bounds in individual cases, the PPN framework offers the advantage of breaking the parameter degeneracy through a combined analysis of $\beta$ and $\gamma$. Such a joint multi-parameter exploration is currently lacking in the context of GW constraints.
}

\section{Summary and outlook}\label{sec:end}

In this work, we explore the quadratic gravity via the post-Newtonian approximation. 
The existence of massive modes in quadratic gravity precludes a standard PPN expansion, meaning that the PN approximation results must be formulated through newly defined potentials.
For a general perfect fluid source, we have rigorously calculated the exact form of its PN expansion up to 1.5PN order in Sec.~\ref{sec:PN}.
The 1PN-order results are shown in \eq{eq:solO2}, with two new potentials $U_R$ and $U_W$ as \eq{eq:URW}.
Defined as the convolutions of the density with respective Yukawa kernels, these two potentials characterize the contributions from the two massive modes to the metric components.
The 1.5PN-order results are shown in \eq{eq:solO3}, with a newly defined potential $\mathcal{V}_a$ as \eq{eq:newV}.
The potential $\mathcal{V}_a$ is proportional to the convolution between the Weyl mode and the PPN vector potential $V_a$; this manifests the high degree of non-linearity inherent in quadratic gravity.
For an N-body system consisting of point-like sources, the 1PN and 1.5PN results are shown as \eq{eq:Nbd1PN} and \eq{eq:Nbd1.5PN}.
These findings pave the way for further dynamical testing of quadratic gravity, including potential constraints using pulsar timing arrays (PTAs).

Given the complexity of the analytical results, we adopt a static point source as a starting point for exploring 2PN phenomena and derive its corresponding 2PN expressions.
Inspired by Ref.~\cite{Hohmann:2013rba}, we calculate the effective PPN parameters $\gamma(r)$ and $\beta(r)$.
The PPN parameter $\gamma(r)$ is easily obtained from the exact 1PN result for general sources, and the result is
\begin{equation*}
\begin{aligned}
% \frac{G_{\mathrm{eff}}(r)}{G}
G_{\mathrm{eff}}(r)
&=1+\frac{1}{3}e^{-m_R r}-\frac{4}{3}e^{-m_W r},\\
\gamma(r)&=\frac{3-e^{-m_R r}-2e^{-m_W r}}{3+e^{-m_R r}-4e^{-m_W r}}\\
 &\approx 1-\frac{2}{3}e^{-m_R r}+\frac{2}{3}e^{-m_W r}.
% \label{eq:Gga}
\end{aligned}
\end{equation*}
In the limit $m_W\to\infty$, the result reduces to the result in $f(R)$ gravity~\cite{Berry:2011pb}.
The minus sign in front of the Weyl sector in $G_{\mathrm{eff}}(r)$ indicates a ghost contribution, which effectively leads to a repulsive interaction.
Another noteworthy feature is that $G_{\mathrm{eff}}(r)$ approaches zero as $r\to 0$, suggesting a strong suppression of gravitational interaction at very short distances induced by the ghost contribution.
In order to ensure that gravity remains attractive, we require $\gamma(r)>0$.
Taking the limit $r\to 0$ of $\gamma(r)$, we obtain
\begin{equation*}
\lim_{r\to 0}\gamma(r)=\frac{m_R+2m_W}{4 m_W-m_R}>0,
\end{equation*}
and thus we have the relation $m_W>m_R/4$ to ensure that gravity remains attractive.

The PPN parameter $\beta(r)$ is calculated in Sec.~\ref{sec:2PN}.
The full expression is shown as \eq{eq:betafull}, and the second-order approximation from the asymptotic expansion of the full expression at $r\to\infty$ is shown as \eq{eq:betaO2}.
By keeping the $1/\epsilon$ terms throughout the derivation, we observed that the divergences vanish identically in the end. 
This behavior mirrors the cancellation of ultraviolet divergences in loop-level dimensional regularization, thereby verifying the robustness of our calculated 2PN expressions.
Throughout the calculation, individual polynomial terms in $r$ arise from both massless and massive sectors. 
Yet, these terms vanish identically in the final sum, such that the remaining 2PN corrections are entirely characterized by the $e^{-mr}$ suppression. 
Such a behavior aligns with the physical intuition that massive modes should not produce long-range polynomial corrections, thereby strengthening the validity of our derivation.
The first-order approximation of $\beta(r)$ is
\begin{equation*}
\begin{aligned}
\beta (r)
\approx& 1
+\frac{1}{3}m_R r \ln(2 m_R r)e^{-m_R r}
\\&
-\frac{4}{3}m_W r \ln(2 m_W r)e^{-m_W r},
\end{aligned}
\end{equation*}
which is characterized by a leading-order behavior of the form $\mathcal{O}(r \ln (r)e^{-mr})$.
This behavior is to be expected, given that similar results have been manifested in scalar-tensor theories~\cite{Hohmann:2013rba, Hohmann:2017qje}. It likely represents a \emph{generic property} of $\beta(r)$ within the framework of modified gravity theories with massive degrees of freedom.

Given the effective $\beta(r)$ and $\gamma(r)$, in Sec.~\ref{sec:exp}, we perform a solar system test to constrain the parameters of quadratic gravity.
The combined constraints from the perihelion precession of Mercury and Cassini radio tracking yield a limit of $m > 23~\mathrm{AU}^{-1}\simeq 1.5\times 10^{-10}{~\rm m}$, where $m$ stands for $m_R$ and $m_W$.
While $f(R)$ gravity is constrained to $m_R > 1700~\mathrm{AU}^{-1}$, the bounds on quadratic gravity are far less restrictive. 
This stems from the fact that $\gamma(r) \equiv 1$ when $m_R = m_W$, which leads to a complete degeneracy in the solar system tests.
Furthermore, we demonstrate that the constraints obtained using the full expression of $\beta(r)$ show little deviation from those derived using only the first-order approximation. This indicates that the first-order approximation of $\beta(r)$ is sufficiently accurate for the purpose of performing constraints.
Parameterized in terms of $\lambda$ and $\mu$, the resulting constraints are $\lambda \lesssim 2.1\times 10^{19}~\mathrm{m}^2$ and $\mu\lesssim 7.1\times 10^{18}~\mathrm{m}^2$.

While the resulting bounds appear loose compared to other established theories, they provide the inaugural valid limits for quadratic gravity in this context, filling a gap in the current literature.
For the future tests, in the weak-field regime, the most precise astronomical tests are currently provided by PTAs. 
The PPN framework established in this work provides a solid theoretical foundation for such PTA-based tests, paving the way for more stringent constraints on quadratic gravity in the near future.

\del{Another robust approach to constraining quadratic gravity is through gravitational waves (GWs). The presence of two additional massive modes leads to extra energy loss during the inspiral phase, which in turn introduces corrections to the GW waveforms}
% ~\cite{Yagi:2011xp, Alves:2022yea, Alves:2025qcx}. 
\del{
% Unlike the PPN framework, where these modes contribute with opposite signs, their contributions to the GW energy flux are additive. 
% This crucial difference allows GW observations to break the $m_R = m_W$ degeneracy encountered in solar system tests. 
In fact, preliminary conclusions have already been drawn from projected constraints using gravitational-wave observations. For the case of GR supplemented by a massive scalar mode, Ref.~\cite{Alves:2025qcx} provides a lower bound of $m_R \gtrsim 1.6 \times 10^{-6} \mathrm{ m}^{-1}$. Similarly, for the scenario of GR supplemented by a massive spin-2 mode, the constraint is given as $m_W \gtrsim 6.9 \times 10^{-6} \mathrm{ m}^{-1}$~\cite{Alves:2025qcx}.
These results surpass those derived from Solar System PPN tests in Sec.~\ref{sec:exp}, where the constraints are $m_R \gtrsim 1700 \text{ AU}^{-1} \simeq 1.6 \times 10^{-6} \text{ m}^{-1}$ for the case of GR supplemented by massive scalar sector and $m_W \gtrsim 1200 \text{ AU}^{-1} \simeq 8.0 \times 10^{-9} \text{ m}^{-1}$ for the case of GR supplemented by massive spin-2 sector.
% By utilizing the predicted waveforms from existing literature, we expect to establish more stringent and independent constraints on quadratic gravity with future GW detections.
However, previously derived gravitational-wave waveforms primarily accounted for modifications in energy loss. With the explicit PPN formalisms and the corresponding orbital corrections now available, we expect to develop more accurate waveform templates and establish more stringent constraints.
}

Furthermore, for testing modes characterized by Yukawa-type corrections, the E{\"o}t-Wash experiments~\cite{Adelberger:2009zz, Kapner:2006si, Perivolaropoulos:2016ucs, Lee:2020zjt} currently provide the most precise constraints.
The E{\"o}t-Wash experiments utilize a high-precision torsion balance to measure the gravitational potential energy between two structured test masses at laboratory scales. By rotating a multi-holed attractor beneath a pendulum, the setup can detect minute deviations from the Newtonian potential, specifically targeting short-range Yukawa-type interactions. Operating at sub-millimeter separations, this mechanism provides the most sensitive probe for the massive modes of gravity, where the exponential suppression $e^{-mr}$ is least dominant.
In the context of $f(R)$ gravity,
 % \lh{where $f(R)\propto R^2$}, 
these laboratory-scale tests have established a remarkably stringent lower bound of $\mu\lesssim  4.5\times10^{-10}~\mathrm{m}^2$~\cite{Kapner:2006si, Perivolaropoulos:2016ucs, Lee:2020zjt}.
Within the framework of quadratic gravity, the effective potential from a point mass takes the form
\begin{equation}
V(r)=\frac{GM}{r}\left(1+\frac{1}{3}e^{-m_R r}-\frac{4}{3}e^{-m_W r}\right).
\end{equation}
Unlike $f(R)$ gravity, in the regime $m_R r \ll 1$, $m_W r\ll 1$, the asymptotic behavior of the potential $V$ is given by
\begin{equation*}
V(r)\approx \frac{GM}{3}(-m_R+4m_W) +\frac{GM}{6}(m_R^2-4m_W^2)r+\mathcal{O}(r^2).
\end{equation*}
That is, very differently from GR, at very short length scales, due to the screening effect of the ghost modes, $V$ is no longer inversely proportional to $r$, but instead becomes a polynomial function of $r$.
Our future expectation is to utilize these short-range gravity measurements to provide a much more rigorous test of the theory’s massive modes.

% However, regarding the detection of modes with Yukawa-type couplings, the Eöt-Wash experiments remain the most sensitive, setting the most precise bounds to date.

\section*{Acknowledgements}

We acknowledge the use of the \texttt{xPPN} package \url{https://github.com/xenos1984/xPPN}~\cite{Hohmann:2020muq} in the theoretical calculations. This work was supported in part by the National Natural Science Foundation of China under Grant No.~12547101. HL was also supported by the start-up fund of Chongqing University under No.~0233005203009, and JZ was supported by the start-up fund of Chongqing University under No.~0233005203006.

\appendix

\bibliography{refs}

\begin{thebibliography}{62}%
\makeatletter
\providecommand \@ifxundefined [1]{%
 \@ifx{#1\undefined}
}%
\providecommand \@ifnum [1]{%
 \ifnum #1\expandafter \@firstoftwo
 \else \expandafter \@secondoftwo
 \fi
}%
\providecommand \@ifx [1]{%
 \ifx #1\expandafter \@firstoftwo
 \else \expandafter \@secondoftwo
 \fi
}%
\providecommand \natexlab [1]{#1}%
\providecommand \enquote  [1]{``#1''}%
\providecommand \bibnamefont  [1]{#1}%
\providecommand \bibfnamefont [1]{#1}%
\providecommand \citenamefont [1]{#1}%
\providecommand \href@noop [0]{\@secondoftwo}%
\providecommand \href [0]{\begingroup \@sanitize@url \@href}%
\providecommand \@href[1]{\@@startlink{#1}\@@href}%
\providecommand \@@href[1]{\endgroup#1\@@endlink}%
\providecommand \@sanitize@url [0]{\catcode `\\12\catcode `\$12\catcode
  `\&12\catcode `\#12\catcode `\^12\catcode `\_12\catcode `\%12\relax}%
\providecommand \@@startlink[1]{}%
\providecommand \@@endlink[0]{}%
\providecommand \url  [0]{\begingroup\@sanitize@url \@url }%
\providecommand \@url [1]{\endgroup\@href {#1}{\urlprefix }}%
\providecommand \urlprefix  [0]{URL }%
\providecommand \Eprint [0]{\href }%
\providecommand \doibase [0]{https://doi.org/}%
\providecommand \selectlanguage [0]{\@gobble}%
\providecommand \bibinfo  [0]{\@secondoftwo}%
\providecommand \bibfield  [0]{\@secondoftwo}%
\providecommand \translation [1]{[#1]}%
\providecommand \BibitemOpen [0]{}%
\providecommand \bibitemStop [0]{}%
\providecommand \bibitemNoStop [0]{.\EOS\space}%
\providecommand \EOS [0]{\spacefactor3000\relax}%
\providecommand \BibitemShut  [1]{\csname bibitem#1\endcsname}%
\let\auto@bib@innerbib\@empty
%</preamble>
\bibitem [{\citenamefont {Will}(2014)}]{Will:2014kxa}%
  \BibitemOpen
  \bibfield  {author} {\bibinfo {author} {\bibfnamefont {C.~M.}\ \bibnamefont
  {Will}},\ }\bibfield  {title} {\bibinfo {title} {{The Confrontation between
  General Relativity and Experiment}},\ }\href
  {https://doi.org/10.12942/lrr-2014-4} {\bibfield  {journal} {\bibinfo
  {journal} {Living Rev. Rel.}\ }\textbf {\bibinfo {volume} {17}},\ \bibinfo
  {pages} {4} (\bibinfo {year} {2014})},\ \Eprint
  {https://arxiv.org/abs/1403.7377} {arXiv:1403.7377 [gr-qc]} \BibitemShut
  {NoStop}%
\bibitem [{\citenamefont {Turyshev}(2009)}]{Turyshev:2008ur}%
  \BibitemOpen
  \bibfield  {author} {\bibinfo {author} {\bibfnamefont {S.~G.}\ \bibnamefont
  {Turyshev}},\ }\bibfield  {title} {\bibinfo {title} {{Experimental Tests of
  General Relativity: Recent Progress and Future Directions}},\ }\href
  {https://doi.org/10.3367/UFNe.0179.200901a.0003} {\bibfield  {journal}
  {\bibinfo  {journal} {Usp. Fiz. Nauk}\ }\textbf {\bibinfo {volume} {179}},\
  \bibinfo {pages} {3034} (\bibinfo {year} {2009})},\ \Eprint
  {https://arxiv.org/abs/0809.3730} {arXiv:0809.3730 [gr-qc]} \BibitemShut
  {NoStop}%
\bibitem [{\citenamefont {Berti}\ \emph {et~al.}(2015)\citenamefont {Berti}
  \emph {et~al.}}]{Berti:2015itd}%
  \BibitemOpen
  \bibfield  {author} {\bibinfo {author} {\bibfnamefont {E.}~\bibnamefont
  {Berti}} \emph {et~al.},\ }\bibfield  {title} {\bibinfo {title} {{Testing
  General Relativity with Present and Future Astrophysical Observations}},\
  }\href {https://doi.org/10.1088/0264-9381/32/24/243001} {\bibfield  {journal}
  {\bibinfo  {journal} {Class. Quant. Grav.}\ }\textbf {\bibinfo {volume}
  {32}},\ \bibinfo {pages} {243001} (\bibinfo {year} {2015})},\ \Eprint
  {https://arxiv.org/abs/1501.07274} {arXiv:1501.07274 [gr-qc]} \BibitemShut
  {NoStop}%
\bibitem [{\citenamefont {Abbott}\ \emph {et~al.}(2016)\citenamefont {Abbott}
  \emph {et~al.}}]{LIGOScientific:2016aoc}%
  \BibitemOpen
  \bibfield  {author} {\bibinfo {author} {\bibfnamefont {B.~P.}\ \bibnamefont
  {Abbott}} \emph {et~al.} (\bibinfo {collaboration} {LIGO Scientific,
  Virgo}),\ }\bibfield  {title} {\bibinfo {title} {{Observation of
  Gravitational Waves from a Binary Black Hole Merger}},\ }\href
  {https://doi.org/10.1103/PhysRevLett.116.061102} {\bibfield  {journal}
  {\bibinfo  {journal} {Phys. Rev. Lett.}\ }\textbf {\bibinfo {volume} {116}},\
  \bibinfo {pages} {061102} (\bibinfo {year} {2016})},\ \Eprint
  {https://arxiv.org/abs/1602.03837} {arXiv:1602.03837 [gr-qc]} \BibitemShut
  {NoStop}%
\bibitem [{\citenamefont {Akiyama}\ \emph {et~al.}(2019)\citenamefont {Akiyama}
  \emph {et~al.}}]{EventHorizonTelescope:2019dse}%
  \BibitemOpen
  \bibfield  {author} {\bibinfo {author} {\bibfnamefont {K.}~\bibnamefont
  {Akiyama}} \emph {et~al.} (\bibinfo {collaboration} {Event Horizon
  Telescope}),\ }\bibfield  {title} {\bibinfo {title} {{First M87 Event Horizon
  Telescope Results. I. The Shadow of the Supermassive Black Hole}},\ }\href
  {https://doi.org/10.3847/2041-8213/ab0ec7} {\bibfield  {journal} {\bibinfo
  {journal} {Astrophys. J. Lett.}\ }\textbf {\bibinfo {volume} {875}},\
  \bibinfo {pages} {L1} (\bibinfo {year} {2019})},\ \Eprint
  {https://arxiv.org/abs/1906.11238} {arXiv:1906.11238 [astro-ph.GA]}
  \BibitemShut {NoStop}%
\bibitem [{\citenamefont {Stelle}(1977)}]{Stelle:1976gc}%
  \BibitemOpen
  \bibfield  {author} {\bibinfo {author} {\bibfnamefont {K.~S.}\ \bibnamefont
  {Stelle}},\ }\bibfield  {title} {\bibinfo {title} {{Renormalization of Higher
  Derivative Quantum Gravity}},\ }\href
  {https://doi.org/10.1103/PhysRevD.16.953} {\bibfield  {journal} {\bibinfo
  {journal} {Phys. Rev. D}\ }\textbf {\bibinfo {volume} {16}},\ \bibinfo
  {pages} {953} (\bibinfo {year} {1977})}\BibitemShut {NoStop}%
\bibitem [{\citenamefont {Stelle}(1978)}]{Stelle:1977ry}%
  \BibitemOpen
  \bibfield  {author} {\bibinfo {author} {\bibfnamefont {K.~S.}\ \bibnamefont
  {Stelle}},\ }\bibfield  {title} {\bibinfo {title} {{Classical Gravity with
  Higher Derivatives}},\ }\href {https://doi.org/10.1007/BF00760427} {\bibfield
   {journal} {\bibinfo  {journal} {Gen. Rel. Grav.}\ }\textbf {\bibinfo
  {volume} {9}},\ \bibinfo {pages} {353} (\bibinfo {year} {1978})}\BibitemShut
  {NoStop}%
\bibitem [{\citenamefont {Buchbinder}\ \emph {et~al.}(2017)\citenamefont
  {Buchbinder}, \citenamefont {Odintsov},\ and\ \citenamefont
  {Shapiro}}]{Buchbinder:1992gdx}%
  \BibitemOpen
  \bibfield  {author} {\bibinfo {author} {\bibfnamefont {I.~L.}\ \bibnamefont
  {Buchbinder}}, \bibinfo {author} {\bibfnamefont {S.~D.}\ \bibnamefont
  {Odintsov}},\ and\ \bibinfo {author} {\bibfnamefont {I.~L.}\ \bibnamefont
  {Shapiro}},\ }\href {https://doi.org/10.1201/9780203758922} {\emph {\bibinfo
  {title} {{Effective Action in Quantum Gravity}}}}\ (\bibinfo  {publisher}
  {Routledge},\ \bibinfo {year} {2017})\BibitemShut {NoStop}%
\bibitem [{\citenamefont {Donoghue}(1994)}]{Donoghue:1994dn}%
  \BibitemOpen
  \bibfield  {author} {\bibinfo {author} {\bibfnamefont {J.~F.}\ \bibnamefont
  {Donoghue}},\ }\bibfield  {title} {\bibinfo {title} {{General relativity as
  an effective field theory: The leading quantum corrections}},\ }\href
  {https://doi.org/10.1103/PhysRevD.50.3874} {\bibfield  {journal} {\bibinfo
  {journal} {Phys. Rev. D}\ }\textbf {\bibinfo {volume} {50}},\ \bibinfo
  {pages} {3874} (\bibinfo {year} {1994})},\ \Eprint
  {https://arxiv.org/abs/gr-qc/9405057} {arXiv:gr-qc/9405057} \BibitemShut
  {NoStop}%
\bibitem [{\citenamefont {Callan}\ \emph {et~al.}(1985)\citenamefont {Callan},
  \citenamefont {Martinec}, \citenamefont {Perry},\ and\ \citenamefont
  {Friedan}}]{Callan:1985ia}%
  \BibitemOpen
  \bibfield  {author} {\bibinfo {author} {\bibfnamefont {C.~G.}\ \bibnamefont
  {Callan}, \bibfnamefont {Jr.}}, \bibinfo {author} {\bibfnamefont {E.~J.}\
  \bibnamefont {Martinec}}, \bibinfo {author} {\bibfnamefont {M.~J.}\
  \bibnamefont {Perry}},\ and\ \bibinfo {author} {\bibfnamefont
  {D.}~\bibnamefont {Friedan}},\ }\bibfield  {title} {\bibinfo {title}
  {{Strings in Background Fields}},\ }\href
  {https://doi.org/10.1016/0550-3213(85)90506-1} {\bibfield  {journal}
  {\bibinfo  {journal} {Nucl. Phys. B}\ }\textbf {\bibinfo {volume} {262}},\
  \bibinfo {pages} {593} (\bibinfo {year} {1985})}\BibitemShut {NoStop}%
\bibitem [{\citenamefont {Gross}\ and\ \citenamefont
  {Sloan}(1987)}]{Gross:1986mw}%
  \BibitemOpen
  \bibfield  {author} {\bibinfo {author} {\bibfnamefont {D.~J.}\ \bibnamefont
  {Gross}}\ and\ \bibinfo {author} {\bibfnamefont {J.~H.}\ \bibnamefont
  {Sloan}},\ }\bibfield  {title} {\bibinfo {title} {{The Quartic Effective
  Action for the Heterotic String}},\ }\href
  {https://doi.org/10.1016/0550-3213(87)90465-2} {\bibfield  {journal}
  {\bibinfo  {journal} {Nucl. Phys. B}\ }\textbf {\bibinfo {volume} {291}},\
  \bibinfo {pages} {41} (\bibinfo {year} {1987})}\BibitemShut {NoStop}%
\bibitem [{\citenamefont {Ashtekar}\ and\ \citenamefont
  {Singh}(2011)}]{Ashtekar:2011ni}%
  \BibitemOpen
  \bibfield  {author} {\bibinfo {author} {\bibfnamefont {A.}~\bibnamefont
  {Ashtekar}}\ and\ \bibinfo {author} {\bibfnamefont {P.}~\bibnamefont
  {Singh}},\ }\bibfield  {title} {\bibinfo {title} {{Loop Quantum Cosmology: A
  Status Report}},\ }\href {https://doi.org/10.1088/0264-9381/28/21/213001}
  {\bibfield  {journal} {\bibinfo  {journal} {Class. Quant. Grav.}\ }\textbf
  {\bibinfo {volume} {28}},\ \bibinfo {pages} {213001} (\bibinfo {year}
  {2011})},\ \Eprint {https://arxiv.org/abs/1108.0893} {arXiv:1108.0893
  [gr-qc]} \BibitemShut {NoStop}%
\bibitem [{\citenamefont {Starobinsky}(1980)}]{Starobinsky:1980te}%
  \BibitemOpen
  \bibfield  {author} {\bibinfo {author} {\bibfnamefont {A.~A.}\ \bibnamefont
  {Starobinsky}},\ }\bibfield  {title} {\bibinfo {title} {{A New Type of
  Isotropic Cosmological Models Without Singularity}},\ }\href
  {https://doi.org/10.1016/0370-2693(80)90670-X} {\bibfield  {journal}
  {\bibinfo  {journal} {Phys. Lett. B}\ }\textbf {\bibinfo {volume} {91}},\
  \bibinfo {pages} {99} (\bibinfo {year} {1980})}\BibitemShut {NoStop}%
\bibitem [{\citenamefont {Mukhanov}(1989)}]{Mukhanov:1989rq}%
  \BibitemOpen
  \bibfield  {author} {\bibinfo {author} {\bibfnamefont {V.~F.}\ \bibnamefont
  {Mukhanov}},\ }\bibfield  {title} {\bibinfo {title} {{Quantum Theory of
  Cosmological Perturbations in R(2) Gravity}},\ }\href
  {https://doi.org/10.1016/0370-2693(89)90467-X} {\bibfield  {journal}
  {\bibinfo  {journal} {Phys. Lett. B}\ }\textbf {\bibinfo {volume} {218}},\
  \bibinfo {pages} {17} (\bibinfo {year} {1989})}\BibitemShut {NoStop}%
\bibitem [{\citenamefont {Whitt}(1984)}]{Whitt:1984pd}%
  \BibitemOpen
  \bibfield  {author} {\bibinfo {author} {\bibfnamefont {B.}~\bibnamefont
  {Whitt}},\ }\bibfield  {title} {\bibinfo {title} {{Fourth Order Gravity as
  General Relativity Plus Matter}},\ }\href
  {https://doi.org/10.1016/0370-2693(84)90332-0} {\bibfield  {journal}
  {\bibinfo  {journal} {Phys. Lett. B}\ }\textbf {\bibinfo {volume} {145}},\
  \bibinfo {pages} {176} (\bibinfo {year} {1984})}\BibitemShut {NoStop}%
\bibitem [{\citenamefont {Myrzakulov}\ \emph {et~al.}(2015)\citenamefont
  {Myrzakulov}, \citenamefont {Odintsov},\ and\ \citenamefont
  {Sebastiani}}]{Myrzakulov:2014hca}%
  \BibitemOpen
  \bibfield  {author} {\bibinfo {author} {\bibfnamefont {R.}~\bibnamefont
  {Myrzakulov}}, \bibinfo {author} {\bibfnamefont {S.}~\bibnamefont
  {Odintsov}},\ and\ \bibinfo {author} {\bibfnamefont {L.}~\bibnamefont
  {Sebastiani}},\ }\bibfield  {title} {\bibinfo {title} {{Inflationary universe
  from higher-derivative quantum gravity}},\ }\href
  {https://doi.org/10.1103/PhysRevD.91.083529} {\bibfield  {journal} {\bibinfo
  {journal} {Phys. Rev. D}\ }\textbf {\bibinfo {volume} {91}},\ \bibinfo
  {pages} {083529} (\bibinfo {year} {2015})},\ \Eprint
  {https://arxiv.org/abs/1412.1073} {arXiv:1412.1073 [gr-qc]} \BibitemShut
  {NoStop}%
\bibitem [{\citenamefont {Elizalde}\ \emph {et~al.}(2017)\citenamefont
  {Elizalde}, \citenamefont {Odintsov}, \citenamefont {Sebastiani},\ and\
  \citenamefont {Myrzakulov}}]{Elizalde:2017mrn}%
  \BibitemOpen
  \bibfield  {author} {\bibinfo {author} {\bibfnamefont {E.}~\bibnamefont
  {Elizalde}}, \bibinfo {author} {\bibfnamefont {S.~D.}\ \bibnamefont
  {Odintsov}}, \bibinfo {author} {\bibfnamefont {L.}~\bibnamefont
  {Sebastiani}},\ and\ \bibinfo {author} {\bibfnamefont {R.}~\bibnamefont
  {Myrzakulov}},\ }\bibfield  {title} {\bibinfo {title} {{Beyond-one-loop
  quantum gravity action yielding both inflation and late-time acceleration}},\
  }\href {https://doi.org/10.1016/j.nuclphysb.2017.06.003} {\bibfield
  {journal} {\bibinfo  {journal} {Nucl. Phys. B}\ }\textbf {\bibinfo {volume}
  {921}},\ \bibinfo {pages} {411} (\bibinfo {year} {2017})},\ \Eprint
  {https://arxiv.org/abs/1706.01879} {arXiv:1706.01879 [gr-qc]} \BibitemShut
  {NoStop}%
\bibitem [{\citenamefont {Akrami}\ \emph {et~al.}(2020)\citenamefont {Akrami}
  \emph {et~al.}}]{Planck:2018jri}%
  \BibitemOpen
  \bibfield  {author} {\bibinfo {author} {\bibfnamefont {Y.}~\bibnamefont
  {Akrami}} \emph {et~al.} (\bibinfo {collaboration} {Planck}),\ }\bibfield
  {title} {\bibinfo {title} {{Planck 2018 results. X. Constraints on
  inflation}},\ }\href {https://doi.org/10.1051/0004-6361/201833887} {\bibfield
   {journal} {\bibinfo  {journal} {Astron. Astrophys.}\ }\textbf {\bibinfo
  {volume} {641}},\ \bibinfo {pages} {A10} (\bibinfo {year} {2020})},\ \Eprint
  {https://arxiv.org/abs/1807.06211} {arXiv:1807.06211 [astro-ph.CO]}
  \BibitemShut {NoStop}%
\bibitem [{\citenamefont {Kanti}\ \emph {et~al.}(1999)\citenamefont {Kanti},
  \citenamefont {Rizos},\ and\ \citenamefont {Tamvakis}}]{Kanti:1998jd}%
  \BibitemOpen
  \bibfield  {author} {\bibinfo {author} {\bibfnamefont {P.}~\bibnamefont
  {Kanti}}, \bibinfo {author} {\bibfnamefont {J.}~\bibnamefont {Rizos}},\ and\
  \bibinfo {author} {\bibfnamefont {K.}~\bibnamefont {Tamvakis}},\ }\bibfield
  {title} {\bibinfo {title} {{Singularity free cosmological solutions in
  quadratic gravity}},\ }\href {https://doi.org/10.1103/PhysRevD.59.083512}
  {\bibfield  {journal} {\bibinfo  {journal} {Phys. Rev. D}\ }\textbf {\bibinfo
  {volume} {59}},\ \bibinfo {pages} {083512} (\bibinfo {year} {1999})},\
  \Eprint {https://arxiv.org/abs/gr-qc/9806085} {arXiv:gr-qc/9806085}
  \BibitemShut {NoStop}%
\bibitem [{\citenamefont {Asorey}\ \emph {et~al.}(2025)\citenamefont {Asorey},
  \citenamefont {Ezquerro},\ and\ \citenamefont {Pardina}}]{Asorey:2024oxw}%
  \BibitemOpen
  \bibfield  {author} {\bibinfo {author} {\bibfnamefont {M.}~\bibnamefont
  {Asorey}}, \bibinfo {author} {\bibfnamefont {F.}~\bibnamefont {Ezquerro}},\
  and\ \bibinfo {author} {\bibfnamefont {M.}~\bibnamefont {Pardina}},\
  }\bibfield  {title} {\bibinfo {title} {{Stability of cosmological
  singularity-free solutions in quadratic gravity}},\ }\href
  {https://doi.org/10.1103/PhysRevD.111.064020} {\bibfield  {journal} {\bibinfo
   {journal} {Phys. Rev. D}\ }\textbf {\bibinfo {volume} {111}},\ \bibinfo
  {pages} {064020} (\bibinfo {year} {2025})},\ \Eprint
  {https://arxiv.org/abs/2412.10111} {arXiv:2412.10111 [gr-qc]} \BibitemShut
  {NoStop}%
\bibitem [{\citenamefont {Duplessis}\ and\ \citenamefont
  {Easson}(2015)}]{Duplessis:2015xva}%
  \BibitemOpen
  \bibfield  {author} {\bibinfo {author} {\bibfnamefont {F.}~\bibnamefont
  {Duplessis}}\ and\ \bibinfo {author} {\bibfnamefont {D.~A.}\ \bibnamefont
  {Easson}},\ }\bibfield  {title} {\bibinfo {title} {{Traversable wormholes and
  non-singular black holes from the vacuum of quadratic gravity}},\ }\href
  {https://doi.org/10.1103/PhysRevD.92.043516} {\bibfield  {journal} {\bibinfo
  {journal} {Phys. Rev. D}\ }\textbf {\bibinfo {volume} {92}},\ \bibinfo
  {pages} {043516} (\bibinfo {year} {2015})},\ \Eprint
  {https://arxiv.org/abs/1506.00988} {arXiv:1506.00988 [gr-qc]} \BibitemShut
  {NoStop}%
\bibitem [{\citenamefont {Berej}\ \emph {et~al.}(2006)\citenamefont {Berej},
  \citenamefont {Matyjasek}, \citenamefont {Tryniecki},\ and\ \citenamefont
  {Woronowicz}}]{Berej:2006cc}%
  \BibitemOpen
  \bibfield  {author} {\bibinfo {author} {\bibfnamefont {W.}~\bibnamefont
  {Berej}}, \bibinfo {author} {\bibfnamefont {J.}~\bibnamefont {Matyjasek}},
  \bibinfo {author} {\bibfnamefont {D.}~\bibnamefont {Tryniecki}},\ and\
  \bibinfo {author} {\bibfnamefont {M.}~\bibnamefont {Woronowicz}},\ }\bibfield
   {title} {\bibinfo {title} {{Regular black holes in quadratic gravity}},\
  }\href {https://doi.org/10.1007/s10714-006-0270-9} {\bibfield  {journal}
  {\bibinfo  {journal} {Gen. Rel. Grav.}\ }\textbf {\bibinfo {volume} {38}},\
  \bibinfo {pages} {885} (\bibinfo {year} {2006})},\ \Eprint
  {https://arxiv.org/abs/hep-th/0606185} {arXiv:hep-th/0606185} \BibitemShut
  {NoStop}%
\bibitem [{\citenamefont {Nojiri}\ \emph {et~al.}(2005)\citenamefont {Nojiri},
  \citenamefont {Odintsov},\ and\ \citenamefont {Tsujikawa}}]{Nojiri:2005sx}%
  \BibitemOpen
  \bibfield  {author} {\bibinfo {author} {\bibfnamefont {S.}~\bibnamefont
  {Nojiri}}, \bibinfo {author} {\bibfnamefont {S.~D.}\ \bibnamefont
  {Odintsov}},\ and\ \bibinfo {author} {\bibfnamefont {S.}~\bibnamefont
  {Tsujikawa}},\ }\bibfield  {title} {\bibinfo {title} {{Properties of
  singularities in (phantom) dark energy universe}},\ }\href
  {https://doi.org/10.1103/PhysRevD.71.063004} {\bibfield  {journal} {\bibinfo
  {journal} {Phys. Rev. D}\ }\textbf {\bibinfo {volume} {71}},\ \bibinfo
  {pages} {063004} (\bibinfo {year} {2005})},\ \Eprint
  {https://arxiv.org/abs/hep-th/0501025} {arXiv:hep-th/0501025} \BibitemShut
  {NoStop}%
\bibitem [{\citenamefont {Bamba}\ \emph {et~al.}(2010)\citenamefont {Bamba},
  \citenamefont {Odintsov}, \citenamefont {Sebastiani},\ and\ \citenamefont
  {Zerbini}}]{Bamba:2010wfw}%
  \BibitemOpen
  \bibfield  {author} {\bibinfo {author} {\bibfnamefont {K.}~\bibnamefont
  {Bamba}}, \bibinfo {author} {\bibfnamefont {S.~D.}\ \bibnamefont {Odintsov}},
  \bibinfo {author} {\bibfnamefont {L.}~\bibnamefont {Sebastiani}},\ and\
  \bibinfo {author} {\bibfnamefont {S.}~\bibnamefont {Zerbini}},\ }\bibfield
  {title} {\bibinfo {title} {{Finite-time future singularities in modified
  Gauss-Bonnet and F(R,G) gravity and singularity avoidance}},\ }\href
  {https://doi.org/10.1140/epjc/s10052-010-1292-8} {\bibfield  {journal}
  {\bibinfo  {journal} {Eur. Phys. J. C}\ }\textbf {\bibinfo {volume} {67}},\
  \bibinfo {pages} {295} (\bibinfo {year} {2010})},\ \Eprint
  {https://arxiv.org/abs/0911.4390} {arXiv:0911.4390 [hep-th]} \BibitemShut
  {NoStop}%
\bibitem [{\citenamefont {Bamba}\ \emph {et~al.}(2008)\citenamefont {Bamba},
  \citenamefont {Nojiri},\ and\ \citenamefont {Odintsov}}]{Bamba:2008ut}%
  \BibitemOpen
  \bibfield  {author} {\bibinfo {author} {\bibfnamefont {K.}~\bibnamefont
  {Bamba}}, \bibinfo {author} {\bibfnamefont {S.}~\bibnamefont {Nojiri}},\ and\
  \bibinfo {author} {\bibfnamefont {S.~D.}\ \bibnamefont {Odintsov}},\
  }\bibfield  {title} {\bibinfo {title} {{The Universe future in modified
  gravity theories: Approaching the finite-time future singularity}},\ }\href
  {https://doi.org/10.1088/1475-7516/2008/10/045} {\bibfield  {journal}
  {\bibinfo  {journal} {JCAP}\ }\textbf {\bibinfo {volume} {10}},\ \bibinfo
  {pages} {045}},\ \Eprint {https://arxiv.org/abs/0807.2575} {arXiv:0807.2575
  [hep-th]} \BibitemShut {NoStop}%
\bibitem [{\citenamefont {Edelstein}\ \emph {et~al.}(2021)\citenamefont
  {Edelstein}, \citenamefont {Ghosh}, \citenamefont {Laddha},\ and\
  \citenamefont {Sarkar}}]{Edelstein:2021jyu}%
  \BibitemOpen
  \bibfield  {author} {\bibinfo {author} {\bibfnamefont {J.~D.}\ \bibnamefont
  {Edelstein}}, \bibinfo {author} {\bibfnamefont {R.}~\bibnamefont {Ghosh}},
  \bibinfo {author} {\bibfnamefont {A.}~\bibnamefont {Laddha}},\ and\ \bibinfo
  {author} {\bibfnamefont {S.}~\bibnamefont {Sarkar}},\ }\bibfield  {title}
  {\bibinfo {title} {{Causality constraints in Quadratic Gravity}},\ }\href
  {https://doi.org/10.1007/JHEP09(2021)150} {\bibfield  {journal} {\bibinfo
  {journal} {JHEP}\ }\textbf {\bibinfo {volume} {09}},\ \bibinfo {pages}
  {150}},\ \Eprint {https://arxiv.org/abs/2107.07424} {arXiv:2107.07424
  [hep-th]} \BibitemShut {NoStop}%
\bibitem [{\citenamefont {Edelstein}\ \emph {et~al.}(2024)\citenamefont
  {Edelstein}, \citenamefont {Ghosh}, \citenamefont {Laddha},\ and\
  \citenamefont {Sarkar}}]{Edelstein:2024jzu}%
  \BibitemOpen
  \bibfield  {author} {\bibinfo {author} {\bibfnamefont {J.~D.}\ \bibnamefont
  {Edelstein}}, \bibinfo {author} {\bibfnamefont {R.}~\bibnamefont {Ghosh}},
  \bibinfo {author} {\bibfnamefont {A.}~\bibnamefont {Laddha}},\ and\ \bibinfo
  {author} {\bibfnamefont {S.}~\bibnamefont {Sarkar}},\ }\bibfield  {title}
  {\bibinfo {title} {{Restoring Causality in Higher Curvature Gravity}},\
  }\href@noop {} {\  (\bibinfo {year} {2024})},\ \Eprint
  {https://arxiv.org/abs/2409.16935} {arXiv:2409.16935 [hep-th]} \BibitemShut
  {NoStop}%
\bibitem [{\citenamefont {Lu}\ \emph {et~al.}(2015)\citenamefont {Lu},
  \citenamefont {Perkins}, \citenamefont {Pope},\ and\ \citenamefont
  {Stelle}}]{Lu:2015cqa}%
  \BibitemOpen
  \bibfield  {author} {\bibinfo {author} {\bibfnamefont {H.}~\bibnamefont
  {Lu}}, \bibinfo {author} {\bibfnamefont {A.}~\bibnamefont {Perkins}},
  \bibinfo {author} {\bibfnamefont {C.~N.}\ \bibnamefont {Pope}},\ and\
  \bibinfo {author} {\bibfnamefont {K.~S.}\ \bibnamefont {Stelle}},\ }\bibfield
   {title} {\bibinfo {title} {{Black Holes in Higher-Derivative Gravity}},\
  }\href {https://doi.org/10.1103/PhysRevLett.114.171601} {\bibfield  {journal}
  {\bibinfo  {journal} {Phys. Rev. Lett.}\ }\textbf {\bibinfo {volume} {114}},\
  \bibinfo {pages} {171601} (\bibinfo {year} {2015})},\ \Eprint
  {https://arxiv.org/abs/1502.01028} {arXiv:1502.01028 [hep-th]} \BibitemShut
  {NoStop}%
\bibitem [{\citenamefont {Giacchini}\ and\ \citenamefont
  {Kol{\'a}{\v{r}}}(2025)}]{Giacchini:2025mlv}%
  \BibitemOpen
  \bibfield  {author} {\bibinfo {author} {\bibfnamefont {B.~L.}\ \bibnamefont
  {Giacchini}}\ and\ \bibinfo {author} {\bibfnamefont {I.}~\bibnamefont
  {Kol{\'a}{\v{r}}}},\ }\href@noop {} {\bibinfo {title} {{Neglected solutions
  in quadratic gravity}}} (\bibinfo {year} {2025}),\ \Eprint
  {https://arxiv.org/abs/2509.07317} {arXiv:2509.07317 [gr-qc]} \BibitemShut
  {NoStop}%
\bibitem [{\citenamefont {Yagi}\ \emph {et~al.}(2012)\citenamefont {Yagi},
  \citenamefont {Stein}, \citenamefont {Yunes},\ and\ \citenamefont
  {Tanaka}}]{Yagi:2011xp}%
  \BibitemOpen
  \bibfield  {author} {\bibinfo {author} {\bibfnamefont {K.}~\bibnamefont
  {Yagi}}, \bibinfo {author} {\bibfnamefont {L.~C.}\ \bibnamefont {Stein}},
  \bibinfo {author} {\bibfnamefont {N.}~\bibnamefont {Yunes}},\ and\ \bibinfo
  {author} {\bibfnamefont {T.}~\bibnamefont {Tanaka}},\ }\bibfield  {title}
  {\bibinfo {title} {{Post-Newtonian, Quasi-Circular Binary Inspirals in
  Quadratic Modified Gravity}},\ }\href
  {https://doi.org/10.1103/PhysRevD.85.064022} {\bibfield  {journal} {\bibinfo
  {journal} {Phys. Rev. D}\ }\textbf {\bibinfo {volume} {85}},\ \bibinfo
  {pages} {064022} (\bibinfo {year} {2012})},\ \bibinfo {note} {[Erratum:
  Phys.Rev.D 93, 029902 (2016)]},\ \Eprint {https://arxiv.org/abs/1110.5950}
  {arXiv:1110.5950 [gr-qc]} \BibitemShut {NoStop}%
\bibitem [{\citenamefont {Alves}\ \emph {et~al.}(2023)\citenamefont {Alves},
  \citenamefont {Reis},\ and\ \citenamefont {Medeiros}}]{Alves:2022yea}%
  \BibitemOpen
  \bibfield  {author} {\bibinfo {author} {\bibfnamefont {M.~F.~S.}\
  \bibnamefont {Alves}}, \bibinfo {author} {\bibfnamefont {L.~F. M. A.~M.}\
  \bibnamefont {Reis}},\ and\ \bibinfo {author} {\bibfnamefont {L.~G.}\
  \bibnamefont {Medeiros}},\ }\bibfield  {title} {\bibinfo {title}
  {{Gravitational waves from inspiraling black holes in quadratic gravity}},\
  }\href {https://doi.org/10.1103/PhysRevD.107.044017} {\bibfield  {journal}
  {\bibinfo  {journal} {Phys. Rev. D}\ }\textbf {\bibinfo {volume} {107}},\
  \bibinfo {pages} {044017} (\bibinfo {year} {2023})},\ \Eprint
  {https://arxiv.org/abs/2206.13672} {arXiv:2206.13672 [gr-qc]} \BibitemShut
  {NoStop}%
\bibitem [{\citenamefont {Alves}\ \emph {et~al.}(2026)\citenamefont {Alves},
  \citenamefont {Medeiros},\ and\ \citenamefont {Rodrigues}}]{Alves:2025qcx}%
  \BibitemOpen
  \bibfield  {author} {\bibinfo {author} {\bibfnamefont {M.~F.~S.}\
  \bibnamefont {Alves}}, \bibinfo {author} {\bibfnamefont {L.~G.}\ \bibnamefont
  {Medeiros}},\ and\ \bibinfo {author} {\bibfnamefont {D.~C.}\ \bibnamefont
  {Rodrigues}},\ }\bibfield  {title} {\bibinfo {title} {{Gravitational
  waveforms from inspiraling compact binaries in quadratic gravity and their
  parametrized post-Einstein characterization}},\ }\href
  {https://doi.org/10.1103/djhs-xvbw} {\bibfield  {journal} {\bibinfo
  {journal} {Phys. Rev. D}\ }\textbf {\bibinfo {volume} {113}},\ \bibinfo
  {pages} {024032} (\bibinfo {year} {2026})},\ \Eprint
  {https://arxiv.org/abs/2507.15571} {arXiv:2507.15571 [gr-qc]} \BibitemShut
  {NoStop}%
\bibitem [{\citenamefont {Nordtvedt}(1968)}]{Nordtvedt:1968qs}%
  \BibitemOpen
  \bibfield  {author} {\bibinfo {author} {\bibfnamefont {K.}~\bibnamefont
  {Nordtvedt}},\ }\bibfield  {title} {\bibinfo {title} {{Equivalence Principle
  for Massive Bodies. 2. Theory}},\ }\href
  {https://doi.org/10.1103/PhysRev.169.1017} {\bibfield  {journal} {\bibinfo
  {journal} {Phys. Rev.}\ }\textbf {\bibinfo {volume} {169}},\ \bibinfo {pages}
  {1017} (\bibinfo {year} {1968})}\BibitemShut {NoStop}%
\bibitem [{\citenamefont {Thorne}\ and\ \citenamefont
  {Will}(1971)}]{Thorne:1971iat}%
  \BibitemOpen
  \bibfield  {author} {\bibinfo {author} {\bibfnamefont {K.~S.}\ \bibnamefont
  {Thorne}}\ and\ \bibinfo {author} {\bibfnamefont {C.~M.}\ \bibnamefont
  {Will}},\ }\bibfield  {title} {\bibinfo {title} {{Theoretical Frameworks for
  Testing Relativistic Gravity. I. Foundations}},\ }\href
  {https://doi.org/10.1086/150803} {\bibfield  {journal} {\bibinfo  {journal}
  {Astrophys. J.}\ }\textbf {\bibinfo {volume} {163}},\ \bibinfo {pages} {595}
  (\bibinfo {year} {1971})}\BibitemShut {NoStop}%
\bibitem [{\citenamefont {Will}(1971{\natexlab{a}})}]{Will:1971zzb}%
  \BibitemOpen
  \bibfield  {author} {\bibinfo {author} {\bibfnamefont {C.~M.}\ \bibnamefont
  {Will}},\ }\bibfield  {title} {\bibinfo {title} {{Theoretical Frameworks for
  Testing Relativistic Gravity. 2. Parametrized Post-Newtonian Hydrodynamics,
  and the Nordtvedt Effect}},\ }\href {https://doi.org/10.1086/150804}
  {\bibfield  {journal} {\bibinfo  {journal} {Astrophys. J.}\ }\textbf
  {\bibinfo {volume} {163}},\ \bibinfo {pages} {611} (\bibinfo {year}
  {1971}{\natexlab{a}})}\BibitemShut {NoStop}%
\bibitem [{\citenamefont {Will}(1971{\natexlab{b}})}]{Will:1971wt}%
  \BibitemOpen
  \bibfield  {author} {\bibinfo {author} {\bibfnamefont {C.~M.}\ \bibnamefont
  {Will}},\ }\bibfield  {title} {\bibinfo {title} {{THEORETICAL FRAMEWORKS FOR
  TESTING RELATIVISTIC GRAVITY. 3. CONSERVATION LAWS, LORENTZ INVARIANCE AND
  VALUES OF THE P P N PARAMETERS}},\ }\href {https://doi.org/10.1086/151124}
  {\bibfield  {journal} {\bibinfo  {journal} {Astrophys. J.}\ }\textbf
  {\bibinfo {volume} {169}},\ \bibinfo {pages} {125} (\bibinfo {year}
  {1971}{\natexlab{b}})}\BibitemShut {NoStop}%
\bibitem [{\citenamefont {Will}(1993)}]{Will_1993}%
  \BibitemOpen
  \bibfield  {author} {\bibinfo {author} {\bibfnamefont {C.~M.}\ \bibnamefont
  {Will}},\ }\href@noop {} {\emph {\bibinfo {title} {Theory and Experiment in
  Gravitational Physics}}}\ (\bibinfo  {publisher} {Cambridge University
  Press},\ \bibinfo {year} {1993})\BibitemShut {NoStop}%
\bibitem [{\citenamefont {Poisson}\ and\ \citenamefont
  {Will}(2014)}]{PoissonWill2014}%
  \BibitemOpen
  \bibfield  {author} {\bibinfo {author} {\bibfnamefont {E.}~\bibnamefont
  {Poisson}}\ and\ \bibinfo {author} {\bibfnamefont {C.~M.}\ \bibnamefont
  {Will}},\ }\href@noop {} {\emph {\bibinfo {title} {Gravity: Newtonian,
  Post-Newtonian, Relativistic}}}\ (\bibinfo  {publisher} {Cambridge University
  Press},\ \bibinfo {address} {Cambridge, UK},\ \bibinfo {year}
  {2014})\BibitemShut {NoStop}%
\bibitem [{\citenamefont {Chiba}\ \emph {et~al.}(2007)\citenamefont {Chiba},
  \citenamefont {Smith},\ and\ \citenamefont {Erickcek}}]{Chiba:2006jp}%
  \BibitemOpen
  \bibfield  {author} {\bibinfo {author} {\bibfnamefont {T.}~\bibnamefont
  {Chiba}}, \bibinfo {author} {\bibfnamefont {T.~L.}\ \bibnamefont {Smith}},\
  and\ \bibinfo {author} {\bibfnamefont {A.~L.}\ \bibnamefont {Erickcek}},\
  }\bibfield  {title} {\bibinfo {title} {{Solar System constraints to general
  f(R) gravity}},\ }\href {https://doi.org/10.1103/PhysRevD.75.124014}
  {\bibfield  {journal} {\bibinfo  {journal} {Phys. Rev. D}\ }\textbf {\bibinfo
  {volume} {75}},\ \bibinfo {pages} {124014} (\bibinfo {year} {2007})},\
  \Eprint {https://arxiv.org/abs/astro-ph/0611867} {arXiv:astro-ph/0611867}
  \BibitemShut {NoStop}%
\bibitem [{\citenamefont {Capozziello}\ and\ \citenamefont
  {Troisi}(2005)}]{Capozziello:2005bu}%
  \BibitemOpen
  \bibfield  {author} {\bibinfo {author} {\bibfnamefont {S.}~\bibnamefont
  {Capozziello}}\ and\ \bibinfo {author} {\bibfnamefont {A.}~\bibnamefont
  {Troisi}},\ }\bibfield  {title} {\bibinfo {title} {{PPN-limit of fourth order
  gravity inspired by scalar-tensor gravity}},\ }\href
  {https://doi.org/10.1103/PhysRevD.72.044022} {\bibfield  {journal} {\bibinfo
  {journal} {Phys. Rev. D}\ }\textbf {\bibinfo {volume} {72}},\ \bibinfo
  {pages} {044022} (\bibinfo {year} {2005})},\ \Eprint
  {https://arxiv.org/abs/astro-ph/0507545} {arXiv:astro-ph/0507545}
  \BibitemShut {NoStop}%
\bibitem [{\citenamefont {Capozziello}\ \emph {et~al.}(2007)\citenamefont
  {Capozziello}, \citenamefont {Stabile},\ and\ \citenamefont
  {Troisi}}]{Capozziello:2007ms}%
  \BibitemOpen
  \bibfield  {author} {\bibinfo {author} {\bibfnamefont {S.}~\bibnamefont
  {Capozziello}}, \bibinfo {author} {\bibfnamefont {A.}~\bibnamefont
  {Stabile}},\ and\ \bibinfo {author} {\bibfnamefont {A.}~\bibnamefont
  {Troisi}},\ }\bibfield  {title} {\bibinfo {title} {{The Newtonian Limit of
  f(R) gravity}},\ }\href {https://doi.org/10.1103/PhysRevD.76.104019}
  {\bibfield  {journal} {\bibinfo  {journal} {Phys. Rev. D}\ }\textbf {\bibinfo
  {volume} {76}},\ \bibinfo {pages} {104019} (\bibinfo {year} {2007})},\
  \Eprint {https://arxiv.org/abs/0708.0723} {arXiv:0708.0723 [gr-qc]}
  \BibitemShut {NoStop}%
\bibitem [{\citenamefont {Capozziello}\ \emph {et~al.}(2009)\citenamefont
  {Capozziello}, \citenamefont {De~Laurentis}, \citenamefont {Nojiri},\ and\
  \citenamefont {Odintsov}}]{Capozziello:2008fn}%
  \BibitemOpen
  \bibfield  {author} {\bibinfo {author} {\bibfnamefont {S.}~\bibnamefont
  {Capozziello}}, \bibinfo {author} {\bibfnamefont {M.}~\bibnamefont
  {De~Laurentis}}, \bibinfo {author} {\bibfnamefont {S.}~\bibnamefont
  {Nojiri}},\ and\ \bibinfo {author} {\bibfnamefont {S.~D.}\ \bibnamefont
  {Odintsov}},\ }\bibfield  {title} {\bibinfo {title} {{f(R) gravity
  constrained by PPN parameters and stochastic background of gravitational
  waves}},\ }\href {https://doi.org/10.1007/s10714-009-0758-1} {\bibfield
  {journal} {\bibinfo  {journal} {Gen. Rel. Grav.}\ }\textbf {\bibinfo {volume}
  {41}},\ \bibinfo {pages} {2313} (\bibinfo {year} {2009})},\ \Eprint
  {https://arxiv.org/abs/0808.1335} {arXiv:0808.1335 [hep-th]} \BibitemShut
  {NoStop}%
\bibitem [{\citenamefont {Qiao}\ \emph {et~al.}(2022)\citenamefont {Qiao},
  \citenamefont {Zhu}, \citenamefont {Li},\ and\ \citenamefont
  {Zhao}}]{Qiao:2021fwi}%
  \BibitemOpen
  \bibfield  {author} {\bibinfo {author} {\bibfnamefont {J.}~\bibnamefont
  {Qiao}}, \bibinfo {author} {\bibfnamefont {T.}~\bibnamefont {Zhu}}, \bibinfo
  {author} {\bibfnamefont {G.}~\bibnamefont {Li}},\ and\ \bibinfo {author}
  {\bibfnamefont {W.}~\bibnamefont {Zhao}},\ }\bibfield  {title} {\bibinfo
  {title} {{Post-Newtonian parameters of ghost-free parity-violating
  gravities}},\ }\href {https://doi.org/10.1088/1475-7516/2022/04/054}
  {\bibfield  {journal} {\bibinfo  {journal} {JCAP}\ }\textbf {\bibinfo
  {volume} {04}}\bibfield  {number} {\bibinfo  {number} { (04)},\ \bibinfo
  {pages} {054}},\ }\Eprint {https://arxiv.org/abs/2110.09033}
  {arXiv:2110.09033 [gr-qc]} \BibitemShut {NoStop}%
\bibitem [{\citenamefont {Toniato}\ and\ \citenamefont
  {Richarte}(2024)}]{Toniato:2024gtx}%
  \BibitemOpen
  \bibfield  {author} {\bibinfo {author} {\bibfnamefont {J.~D.}\ \bibnamefont
  {Toniato}}\ and\ \bibinfo {author} {\bibfnamefont {M.~G.}\ \bibnamefont
  {Richarte}},\ }\bibfield  {title} {\bibinfo {title} {{Post-Newtonian analysis
  of regularized 4D Einstein-Gauss-Bonnet theory: Complete set of PPN
  parameters and observational constraints}},\ }\href
  {https://doi.org/10.1103/PhysRevD.109.104068} {\bibfield  {journal} {\bibinfo
   {journal} {Phys. Rev. D}\ }\textbf {\bibinfo {volume} {109}},\ \bibinfo
  {pages} {104068} (\bibinfo {year} {2024})},\ \Eprint
  {https://arxiv.org/abs/2402.13951} {arXiv:2402.13951 [gr-qc]} \BibitemShut
  {NoStop}%
\bibitem [{\citenamefont {Richarte}\ and\ \citenamefont
  {Toniato}(2025)}]{Richarte:2025dag}%
  \BibitemOpen
  \bibfield  {author} {\bibinfo {author} {\bibfnamefont {M.~G.}\ \bibnamefont
  {Richarte}}\ and\ \bibinfo {author} {\bibfnamefont {J.~D.}\ \bibnamefont
  {Toniato}},\ }\bibfield  {title} {\bibinfo {title} {{Exploring scalarized
  Einstein-Gauss-Bonnet theories through the lens of parametrized
  post-Newtonian formalism}},\ }\href {https://doi.org/10.1103/l23c-kklp}
  {\bibfield  {journal} {\bibinfo  {journal} {Phys. Rev. D}\ }\textbf {\bibinfo
  {volume} {112}},\ \bibinfo {pages} {024019} (\bibinfo {year} {2025})},\
  \Eprint {https://arxiv.org/abs/2503.13339} {arXiv:2503.13339 [gr-qc]}
  \BibitemShut {NoStop}%
\bibitem [{\citenamefont {Vainshtein}(1972)}]{Vainshtein:1972sx}%
  \BibitemOpen
  \bibfield  {author} {\bibinfo {author} {\bibfnamefont {A.~I.}\ \bibnamefont
  {Vainshtein}},\ }\bibfield  {title} {\bibinfo {title} {{To the problem of
  nonvanishing gravitation mass}},\ }\href
  {https://doi.org/10.1016/0370-2693(72)90147-5} {\bibfield  {journal}
  {\bibinfo  {journal} {Phys. Lett. B}\ }\textbf {\bibinfo {volume} {39}},\
  \bibinfo {pages} {393} (\bibinfo {year} {1972})}\BibitemShut {NoStop}%
\bibitem [{\citenamefont {{Avilez-Lopez}}\ \emph {et~al.}(2015)\citenamefont
  {{Avilez-Lopez}}, \citenamefont {Padilla}, \citenamefont {Saffin},\ and\
  \citenamefont {Skordis}}]{Avilez-Lopez:2015dja}%
  \BibitemOpen
  \bibfield  {author} {\bibinfo {author} {\bibfnamefont {A.}~\bibnamefont
  {{Avilez-Lopez}}}, \bibinfo {author} {\bibfnamefont {A.}~\bibnamefont
  {Padilla}}, \bibinfo {author} {\bibfnamefont {P.~M.}\ \bibnamefont
  {Saffin}},\ and\ \bibinfo {author} {\bibfnamefont {C.}~\bibnamefont
  {Skordis}},\ }\bibfield  {title} {\bibinfo {title} {The {{Parametrized
  Post-Newtonian-Vainshteinian}} formalism},\ }\href
  {https://doi.org/10.1088/1475-7516/2015/06/044} {\bibfield  {journal}
  {\bibinfo  {journal} {Journal of Cosmology and Astroparticle Physics}\
  }\textbf {\bibinfo {volume} {06}}\bibfield  {number} {\bibinfo  {number} {
  (06)},\ \bibinfo {pages} {044}},\ }\Eprint {https://arxiv.org/abs/1501.01985}
  {arXiv:1501.01985 [gr-qc]} \BibitemShut {NoStop}%
\bibitem [{\citenamefont {Hohmann}(2021)}]{Hohmann:2020muq}%
  \BibitemOpen
  \bibfield  {author} {\bibinfo {author} {\bibfnamefont {M.}~\bibnamefont
  {Hohmann}},\ }\bibfield  {title} {\bibinfo {title} {{xPPN: an implementation
  of the parametrized post-Newtonian formalism using xAct for Mathematica}},\
  }\href {https://doi.org/10.1140/epjc/s10052-021-09183-9} {\bibfield
  {journal} {\bibinfo  {journal} {Eur. Phys. J. C}\ }\textbf {\bibinfo {volume}
  {81}},\ \bibinfo {pages} {504} (\bibinfo {year} {2021})},\ \Eprint
  {https://arxiv.org/abs/2012.14984} {arXiv:2012.14984 [gr-qc]} \BibitemShut
  {NoStop}%
\bibitem [{\citenamefont {Hohmann}\ \emph {et~al.}(2013)\citenamefont
  {Hohmann}, \citenamefont {Jarv}, \citenamefont {Kuusk},\ and\ \citenamefont
  {Randla}}]{Hohmann:2013rba}%
  \BibitemOpen
  \bibfield  {author} {\bibinfo {author} {\bibfnamefont {M.}~\bibnamefont
  {Hohmann}}, \bibinfo {author} {\bibfnamefont {L.}~\bibnamefont {Jarv}},
  \bibinfo {author} {\bibfnamefont {P.}~\bibnamefont {Kuusk}},\ and\ \bibinfo
  {author} {\bibfnamefont {E.}~\bibnamefont {Randla}},\ }\bibfield  {title}
  {\bibinfo {title} {{Post-Newtonian parameters $\gamma$ and $\beta$ of
  scalar-tensor gravity with a general potential}},\ }\href
  {https://doi.org/10.1103/PhysRevD.88.084054} {\bibfield  {journal} {\bibinfo
  {journal} {Phys. Rev. D}\ }\textbf {\bibinfo {volume} {88}},\ \bibinfo
  {pages} {084054} (\bibinfo {year} {2013})},\ \bibinfo {note} {[Erratum:
  Phys.Rev.D 89, 069901 (2014)]},\ \Eprint {https://arxiv.org/abs/1309.0031}
  {arXiv:1309.0031 [gr-qc]} \BibitemShut {NoStop}%
\bibitem [{\citenamefont {Berry}\ and\ \citenamefont
  {Gair}(2011)}]{Berry:2011pb}%
  \BibitemOpen
  \bibfield  {author} {\bibinfo {author} {\bibfnamefont {C.~P.~L.}\
  \bibnamefont {Berry}}\ and\ \bibinfo {author} {\bibfnamefont {J.~R.}\
  \bibnamefont {Gair}},\ }\bibfield  {title} {\bibinfo {title} {{Linearized
  f(R) Gravity: Gravitational Radiation and Solar System Tests}},\ }\href
  {https://doi.org/10.1103/PhysRevD.83.104022} {\bibfield  {journal} {\bibinfo
  {journal} {Phys. Rev. D}\ }\textbf {\bibinfo {volume} {83}},\ \bibinfo
  {pages} {104022} (\bibinfo {year} {2011})},\ \bibinfo {note} {[Erratum:
  Phys.Rev.D 85, 089906 (2012)]},\ \Eprint {https://arxiv.org/abs/1104.0819}
  {arXiv:1104.0819 [gr-qc]} \BibitemShut {NoStop}%
\bibitem [{\citenamefont {Hohmann}\ and\ \citenamefont
  {Sch{\"a}rer}(2017)}]{Hohmann:2017qje}%
  \BibitemOpen
  \bibfield  {author} {\bibinfo {author} {\bibfnamefont {M.}~\bibnamefont
  {Hohmann}}\ and\ \bibinfo {author} {\bibfnamefont {A.}~\bibnamefont
  {Sch{\"a}rer}},\ }\bibfield  {title} {\bibinfo {title} {{Post-Newtonian
  parameters {\ensuremath{\gamma}} and {\ensuremath{\beta}} of scalar-tensor
  gravity for a homogeneous gravitating sphere}},\ }\href
  {https://doi.org/10.1103/PhysRevD.96.104026} {\bibfield  {journal} {\bibinfo
  {journal} {Phys. Rev. D}\ }\textbf {\bibinfo {volume} {96}},\ \bibinfo
  {pages} {104026} (\bibinfo {year} {2017})},\ \Eprint
  {https://arxiv.org/abs/1708.07851} {arXiv:1708.07851 [gr-qc]} \BibitemShut
  {NoStop}%
\bibitem [{\citenamefont {Giacchini}(2017)}]{Giacchini:2016nta}%
  \BibitemOpen
  \bibfield  {author} {\bibinfo {author} {\bibfnamefont {B.~L.}\ \bibnamefont
  {Giacchini}},\ }\bibfield  {title} {\bibinfo {title} {{Experimental limits on
  the free parameters of higher-derivative gravity}},\ }in\ \href
  {https://doi.org/10.1142/9789813226609_0109} {\emph {\bibinfo {booktitle}
  {{14th Marcel Grossmann Meeting on Recent Developments in Theoretical and
  Experimental General Relativity, Astrophysics, and Relativistic Field
  Theories}}}},\ Vol.~\bibinfo {volume} {2}\ (\bibinfo {year} {2017})\ pp.\
  \bibinfo {pages} {1340--1345},\ \Eprint {https://arxiv.org/abs/1612.01823}
  {arXiv:1612.01823 [gr-qc]} \BibitemShut {NoStop}%
\bibitem [{\citenamefont {Will}(2006)}]{Will:2005va}%
  \BibitemOpen
  \bibfield  {author} {\bibinfo {author} {\bibfnamefont {C.~M.}\ \bibnamefont
  {Will}},\ }\bibfield  {title} {\bibinfo {title} {{The Confrontation between
  general relativity and experiment}},\ }\href
  {https://doi.org/10.12942/lrr-2006-3} {\bibfield  {journal} {\bibinfo
  {journal} {Living Rev. Rel.}\ }\textbf {\bibinfo {volume} {9}},\ \bibinfo
  {pages} {3} (\bibinfo {year} {2006})},\ \Eprint
  {https://arxiv.org/abs/gr-qc/0510072} {arXiv:gr-qc/0510072} \BibitemShut
  {NoStop}%
\bibitem [{\citenamefont {Bertotti}\ \emph {et~al.}(2003)\citenamefont
  {Bertotti}, \citenamefont {Iess},\ and\ \citenamefont
  {Tortora}}]{Bertotti:2003rm}%
  \BibitemOpen
  \bibfield  {author} {\bibinfo {author} {\bibfnamefont {B.}~\bibnamefont
  {Bertotti}}, \bibinfo {author} {\bibfnamefont {L.}~\bibnamefont {Iess}},\
  and\ \bibinfo {author} {\bibfnamefont {P.}~\bibnamefont {Tortora}},\
  }\bibfield  {title} {\bibinfo {title} {{A test of general relativity using
  radio links with the Cassini spacecraft}},\ }\href
  {https://doi.org/10.1038/nature01997} {\bibfield  {journal} {\bibinfo
  {journal} {Nature}\ }\textbf {\bibinfo {volume} {425}},\ \bibinfo {pages}
  {374} (\bibinfo {year} {2003})}\BibitemShut {NoStop}%
\bibitem [{\citenamefont {{Hofmann}}\ \emph {et~al.}(2010)\citenamefont
  {{Hofmann}}, \citenamefont {{M{\"u}ller}},\ and\ \citenamefont
  {{Biskupek}}}]{2010A&A...522L...5H}%
  \BibitemOpen
  \bibfield  {author} {\bibinfo {author} {\bibfnamefont {F.}~\bibnamefont
  {{Hofmann}}}, \bibinfo {author} {\bibfnamefont {J.}~\bibnamefont
  {{M{\"u}ller}}},\ and\ \bibinfo {author} {\bibfnamefont {L.}~\bibnamefont
  {{Biskupek}}},\ }\bibfield  {title} {\bibinfo {title} {{Lunar laser ranging
  test of the Nordtvedt parameter and a possible variation in the gravitational
  constant}},\ }\href {https://doi.org/10.1051/0004-6361/201015659} {\bibfield
  {journal} {\bibinfo  {journal} {Astronomy and Astrophysics}\ }\textbf
  {\bibinfo {volume} {522}},\ \bibinfo {eid} {L5} (\bibinfo {year}
  {2010})}\BibitemShut {NoStop}%
\bibitem [{\citenamefont {Barack}\ \emph {et~al.}(2019)\citenamefont {Barack}
  \emph {et~al.}}]{Barack:2018yly}%
  \BibitemOpen
  \bibfield  {author} {\bibinfo {author} {\bibfnamefont {L.}~\bibnamefont
  {Barack}} \emph {et~al.},\ }\bibfield  {title} {\bibinfo {title} {{Black
  holes, gravitational waves and fundamental physics: a roadmap}},\ }\href
  {https://doi.org/10.1088/1361-6382/ab0587} {\bibfield  {journal} {\bibinfo
  {journal} {Class. Quant. Grav.}\ }\textbf {\bibinfo {volume} {36}},\ \bibinfo
  {pages} {143001} (\bibinfo {year} {2019})},\ \Eprint
  {https://arxiv.org/abs/1806.05195} {arXiv:1806.05195 [gr-qc]} \BibitemShut
  {NoStop}%
\bibitem [{\citenamefont {Berti}\ \emph {et~al.}(2018)\citenamefont {Berti},
  \citenamefont {Yagi},\ and\ \citenamefont {Yunes}}]{Berti:2018cxi}%
  \BibitemOpen
  \bibfield  {author} {\bibinfo {author} {\bibfnamefont {E.}~\bibnamefont
  {Berti}}, \bibinfo {author} {\bibfnamefont {K.}~\bibnamefont {Yagi}},\ and\
  \bibinfo {author} {\bibfnamefont {N.}~\bibnamefont {Yunes}},\ }\bibfield
  {title} {\bibinfo {title} {{Extreme Gravity Tests with Gravitational Waves
  from Compact Binary Coalescences: (I) Inspiral-Merger}},\ }\href
  {https://doi.org/10.1007/s10714-018-2362-8} {\bibfield  {journal} {\bibinfo
  {journal} {Gen. Rel. Grav.}\ }\textbf {\bibinfo {volume} {50}},\ \bibinfo
  {pages} {46} (\bibinfo {year} {2018})},\ \Eprint
  {https://arxiv.org/abs/1801.03208} {arXiv:1801.03208 [gr-qc]} \BibitemShut
  {NoStop}%
\bibitem [{\citenamefont {Alves}\ \emph {et~al.}(2025)\citenamefont {Alves},
  \citenamefont {Cuzinatto}, \citenamefont {de~Melo}, \citenamefont
  {Medeiros},\ and\ \citenamefont {Pompeia}}]{Alves:2024gwi}%
  \BibitemOpen
  \bibfield  {author} {\bibinfo {author} {\bibfnamefont {M.~F.~S.}\
  \bibnamefont {Alves}}, \bibinfo {author} {\bibfnamefont {R.~R.}\ \bibnamefont
  {Cuzinatto}}, \bibinfo {author} {\bibfnamefont {C.~A.~M.}\ \bibnamefont
  {de~Melo}}, \bibinfo {author} {\bibfnamefont {L.~G.}\ \bibnamefont
  {Medeiros}},\ and\ \bibinfo {author} {\bibfnamefont {P.~J.}\ \bibnamefont
  {Pompeia}},\ }\bibfield  {title} {\bibinfo {title} {{Gravitational waves
  emission in quadratic gravity: Longitudinal modes, angular momentum emission,
  and positivity of the radiated power}},\ }\href
  {https://doi.org/10.1103/PhysRevD.111.084055} {\bibfield  {journal} {\bibinfo
   {journal} {Phys. Rev. D}\ }\textbf {\bibinfo {volume} {111}},\ \bibinfo
  {pages} {084055} (\bibinfo {year} {2025})},\ \Eprint
  {https://arxiv.org/abs/2411.10098} {arXiv:2411.10098 [gr-qc]} \BibitemShut
  {NoStop}%
\bibitem [{\citenamefont {Adelberger}\ \emph {et~al.}(2009)\citenamefont
  {Adelberger}, \citenamefont {Gundlach}, \citenamefont {Heckel}, \citenamefont
  {Hoedl},\ and\ \citenamefont {Schlamminger}}]{Adelberger:2009zz}%
  \BibitemOpen
  \bibfield  {author} {\bibinfo {author} {\bibfnamefont {E.~G.}\ \bibnamefont
  {Adelberger}}, \bibinfo {author} {\bibfnamefont {J.~H.}\ \bibnamefont
  {Gundlach}}, \bibinfo {author} {\bibfnamefont {B.~R.}\ \bibnamefont
  {Heckel}}, \bibinfo {author} {\bibfnamefont {S.}~\bibnamefont {Hoedl}},\ and\
  \bibinfo {author} {\bibfnamefont {S.}~\bibnamefont {Schlamminger}},\
  }\bibfield  {title} {\bibinfo {title} {{Torsion balance experiments: A
  low-energy frontier of particle physics}},\ }\href
  {https://doi.org/10.1016/j.ppnp.2008.08.002} {\bibfield  {journal} {\bibinfo
  {journal} {Prog. Part. Nucl. Phys.}\ }\textbf {\bibinfo {volume} {62}},\
  \bibinfo {pages} {102} (\bibinfo {year} {2009})}\BibitemShut {NoStop}%
\bibitem [{\citenamefont {Kapner}\ \emph {et~al.}(2007)\citenamefont {Kapner},
  \citenamefont {Cook}, \citenamefont {Adelberger}, \citenamefont {Gundlach},
  \citenamefont {Heckel}, \citenamefont {Hoyle},\ and\ \citenamefont
  {Swanson}}]{Kapner:2006si}%
  \BibitemOpen
  \bibfield  {author} {\bibinfo {author} {\bibfnamefont {D.~J.}\ \bibnamefont
  {Kapner}}, \bibinfo {author} {\bibfnamefont {T.~S.}\ \bibnamefont {Cook}},
  \bibinfo {author} {\bibfnamefont {E.~G.}\ \bibnamefont {Adelberger}},
  \bibinfo {author} {\bibfnamefont {J.~H.}\ \bibnamefont {Gundlach}}, \bibinfo
  {author} {\bibfnamefont {B.~R.}\ \bibnamefont {Heckel}}, \bibinfo {author}
  {\bibfnamefont {C.~D.}\ \bibnamefont {Hoyle}},\ and\ \bibinfo {author}
  {\bibfnamefont {H.~E.}\ \bibnamefont {Swanson}},\ }\bibfield  {title}
  {\bibinfo {title} {{Tests of the gravitational inverse-square law below the
  dark-energy length scale}},\ }\href
  {https://doi.org/10.1103/PhysRevLett.98.021101} {\bibfield  {journal}
  {\bibinfo  {journal} {Phys. Rev. Lett.}\ }\textbf {\bibinfo {volume} {98}},\
  \bibinfo {pages} {021101} (\bibinfo {year} {2007})},\ \Eprint
  {https://arxiv.org/abs/hep-ph/0611184} {arXiv:hep-ph/0611184} \BibitemShut
  {NoStop}%
\bibitem [{\citenamefont {Perivolaropoulos}(2017)}]{Perivolaropoulos:2016ucs}%
  \BibitemOpen
  \bibfield  {author} {\bibinfo {author} {\bibfnamefont {L.}~\bibnamefont
  {Perivolaropoulos}},\ }\bibfield  {title} {\bibinfo {title} {{Submillimeter
  spatial oscillations of Newton{\textquoteright}s constant: Theoretical models
  and laboratory tests}},\ }\href {https://doi.org/10.1103/PhysRevD.95.084050}
  {\bibfield  {journal} {\bibinfo  {journal} {Phys. Rev. D}\ }\textbf {\bibinfo
  {volume} {95}},\ \bibinfo {pages} {084050} (\bibinfo {year} {2017})},\
  \Eprint {https://arxiv.org/abs/1611.07293} {arXiv:1611.07293 [gr-qc]}
  \BibitemShut {NoStop}%
\bibitem [{\citenamefont {Lee}\ \emph {et~al.}(2020)\citenamefont {Lee},
  \citenamefont {Adelberger}, \citenamefont {Cook}, \citenamefont {Fleischer},\
  and\ \citenamefont {Heckel}}]{Lee:2020zjt}%
  \BibitemOpen
  \bibfield  {author} {\bibinfo {author} {\bibfnamefont {J.~G.}\ \bibnamefont
  {Lee}}, \bibinfo {author} {\bibfnamefont {E.~G.}\ \bibnamefont {Adelberger}},
  \bibinfo {author} {\bibfnamefont {T.~S.}\ \bibnamefont {Cook}}, \bibinfo
  {author} {\bibfnamefont {S.~M.}\ \bibnamefont {Fleischer}},\ and\ \bibinfo
  {author} {\bibfnamefont {B.~R.}\ \bibnamefont {Heckel}},\ }\bibfield  {title}
  {\bibinfo {title} {{New Test of the Gravitational $1/r^2$ Law at Separations
  down to 52 $\mu$m}},\ }\href {https://doi.org/10.1103/PhysRevLett.124.101101}
  {\bibfield  {journal} {\bibinfo  {journal} {Phys. Rev. Lett.}\ }\textbf
  {\bibinfo {volume} {124}},\ \bibinfo {pages} {101101} (\bibinfo {year}
  {2020})},\ \Eprint {https://arxiv.org/abs/2002.11761} {arXiv:2002.11761
  [hep-ex]} \BibitemShut {NoStop}%
\end{thebibliography}%

\end{document}